\documentclass[aps,twocolumn,preprintnumbers,prd,superscriptaddress,nofootinbib,10pt,floatfix]{revtex4-2}
\usepackage[utf8]{inputenc}
\usepackage{graphicx,amssymb,amsmath,amsthm,amsfonts,epsfig,epsf}
\usepackage[usenames,dvipsnames,svgnames,table]{xcolor}
\usepackage{epstopdf}
\usepackage[caption=false]{subfig}
\definecolor{darkred}{rgb}{0.5,0,0}
\usepackage{bm}
\usepackage{dcolumn}
\usepackage{latexsym}
\usepackage{rotating}
\usepackage{longtable}

\usepackage{enumerate}
\usepackage{tensor,multirow}
\usepackage[normalem]{ulem}

\usepackage[english]{babel}

\usepackage{amsmath}
\usepackage{graphicx}
\usepackage[colorlinks=true, allcolors=blue]{hyperref}
\usepackage{scalerel,amssymb}
\usepackage{cancel}
\usepackage{tcolorbox}
\usepackage{empheq}
\usepackage{listings}
\usepackage{xcolor}
\begin{document}
%\title{Binary Proca stars: dynamics and gravitational wave emission from eccentric mergers}
\title{Eccentric mergers of binary Proca stars}

\author{Gabriele Palloni}
\affiliation{Departamento de Astronomía y Astrofísica, Universitat de València, \\
Dr.\ Moliner 50, 46100, Burjassot (Valencia), Spain}

\author{Nicolas Sanchis-Gual}
\affiliation{Departamento de Astronomía y Astrofísica, Universitat de València, \\
Dr.\ Moliner 50, 46100, Burjassot (Valencia), Spain}

\author{José A. Font}
\affiliation{Departamento de Astronomía y Astrofísica, Universitat de València, \\
Dr.\ Moliner 50, 46100, Burjassot (Valencia), Spain}
\affiliation{Observatori Astron\`{o}mic, Universitat de Val\`{e}ncia,
C/ Catedr\'{a}tico Jos\'{e} Beltr\'{a}n 2, 46980, Paterna (Val\`{e}ncia), Spain}

\author{Carlos Herdeiro}
\affiliation{Departamento de Matemática da Universidade de Aveiro \\
and Centre for Research and Development in Mathematics and Applications (CIDMA) \\
Campus de Santiago, 3810-193 Aveiro, Portugal}

\author{Eugen Radu}
\affiliation{Departamento de Matemática da Universidade de Aveiro \\
and Centre for Research and Development in Mathematics and Applications (CIDMA) \\
Campus de Santiago, 3810-193 Aveiro, Portugal}

\begin{abstract}
  %Using numerical relativity simulations we explore the dynamics and gravitational wave emission of binary spinning Proca star mergers in eccentric orbits. We choose three Proca star configurations, with different masses, spins and compactness, and vary some of the parameters of the binary (initial velocity and phase difference). We investigate their role in the outcome of the mergers and the gravitational waveforms. As reported in the head-on case, the phase difference between the stars has a paramount impact on everything. \TF{To be rewritten when paper is finished.}

  We present a numerical relativity study of eccentric mergers of equal-mass rotating $\bar m=1$ Proca stars, focusing on their gravitational-wave (GW) emission. By systematically varying key binary parameters, such as the initial orbital boost, which determines the orbital angular momentum, and the relative phase between the stars, we examine how the internal phase structure of the Proca field influences the merger dynamics and the properties of the emitted GWs. Our simulations demonstrate that the relative phase has paramount impact on the post-merger evolution, resulting in prompt black hole formation accompanied by a transient Proca remnant, the formation of a hypermassive $\bar m=1$ Proca star or even the emergence of a dynamically-unstable spinning $\bar m=2$ Proca star. Under certain conditions, the GW signal exhibits significant odd-modes (e.g., the $\ell=m=3$ mode) that are absent in conventional black hole mergers, potentially serving as unique signatures of these exotic objects. Our findings offer new insights into the phenomenology of bosonic star mergers and the potential astrophysical role of ultralight bosonic fields.
\end{abstract}
\maketitle

%%%%%%%%%%%%%%%%%%%%%%
\section{Introduction}
%%%%%%%%%%%%%%%%%%%%%%

Gravitational-wave (GW) observations have significantly advanced our understanding of stellar-origin compact objects. To date, the LIGO-Virgo-KAGRA (LVK) Collaboration has reported nearly one hundred signals in previous Observing runs (O1-O3) and two hundred more candidate events in O4 from compact binary coalescences, including binary black hole (BBH) mergers, binary neutron star (BNS) mergers, and mixed black hole–neutron star systems~\cite{GWTC1,GWTC2,GWTC2.1,GWTC3}. These detections have revealed a population of black holes spanning from stellar to intermediate masses, enabling in-depth analyses of their mass and spin distributions~\cite{abbott2019binary,abbott2022search,abbott2023population}. As detector sensitivities improve, the fundamental properties of compact objects will be probed with even greater precision. Future observation runs and next-generation detectors, such as the Einstein Telescope~\cite{Punturo:2010,abac2025science} and Cosmic Explorer~\cite{reitze2019cosmic,Evans:2021}, promise to deliver studies of unprecedented detail, which will be essential for testing the Kerr black hole paradigm and constraining the equation of state of dense matter inside neutron stars.

Although compact objects of stellar origin constitute the main target of current GW detectors, current technology could in principle observe the ``unexpected", i.e.~new, theoretically hypothesized, phenomena, provided the frequency range of the associated emission falls within the sensitivity range. In analogy to traditional astronomy, where great discoveries have ensued after the opening of each new window in the electromagnetic spectrum (X-rays, radio waves, gamma rays), one might expect GWs to help find new exotic objects lurking in the Universe~\cite{bezares2024exotic}. Since those yet unknown hypothetical objects have not been discovered through electromagnetic observations, it could be argued that such entities only interact gravitationally with visible matter, making them candidates for dark matter.

In this context, considerable attention has focused on ultralight bosonic fields, which have been proposed as candidates for a fraction of the dark matter in galaxies~\cite{veltmaat2018formation,ferreira2021ultra,matos2024short}. These hypothetical particles have masses in the range of $10^{-10}$ to $10^{-22}$ eV, with small or vanishing couplings to the visible standard model sector, rendering them virtually undetectable in conventional particle accelerator experiments. However, they exhibit intriguing properties, notably, they can induce a superradiant instability around Kerr black holes, effectively extracting energy and angular momentum and causing black holes to spin down~\cite{cardoso2018constraining,east2017superradiant}. In the case of a complex bosonic field, this process may lead to the formation of a Kerr black hole endowed with bosonic hair~\cite{Herdeiro:2014goa,Herdeiro:2016tmi}. In some scenarios, once such a bosonic cloud is established around the black hole, it could gradually dissipate by emitting continuous gravitational waves. Despite several studies that analyze LIGO and Virgo data in search of these signals, so far no evidence of such emission has been detected~\cite{LVK-boson-clouds}. 

Another compelling possibility that links ultralight bosonic fields with exotic compact objects is that these particles can clump together to form horizonless, localized, and self-gravitating solitons known as bosonic stars, characterized by a specific eigenfrequency, $\omega/\mu$. These solitonic objects are solutions of the Einstein–(complex, massive) Klein–Gordon system in the case of scalar boson stars~\cite{Schunck:2003kk} and of the Einstein–(complex) Proca system for vector boson stars (or Proca stars)~\cite{BRITO2016291}. Within the considered range for the particle mass $\mu$, these stars can have masses spanning from $1$ to $10^{12}$ $M_{\odot}$ and, in some regions of the parameter space be at least as compact as neutron stars, making them viable astrophysical compact objects. Moreover, if bosonic stars exist in nature, they may form binaries that would eventually merge and emit distinctive GW signals. Detecting such signals could provide evidence for the existence of these exotic objects, as suggested by studies of GW events such as GW190521 and other high-mass mergers~\cite{PhysRevLett.126.081101,Bustillo:2023,luna2024numerical}. 

To confidently claim a GW detection from the merger of exotic compact objects, waveform models are needed to match-filter the data collected by the detectors against template banks~\cite{evstafyeva2024gravitational}. Obtaining accurate waveforms through numerical relativity simulations of boson star collisions is not an easy task~\cite{palenzuela2017gravitational,PhysRevD.106.124011,ge2024hair}. Early attempts assumed approximations for the initial data~\cite{PhysRevD.77.044036,Bezares_2018,PhysRevD.99.024017,jaramillo2022head,helfer2022malaise,croft2023gravitational} and only recently constraint-satisfying initial data have been obtained and evolved for scalar boson stars~\cite{aurrekoetxea2023cttk,Siemonsen:2023a,Siemonsen:2023b,atteneder2024boson,aurrekoetxea2025grtresna}. In addition, bosonic stars have specific properties that distinguish them from black holes and neutron stars. On the one hand, the angular momentum of the rotating solution is quantized, for fixed number of bosonic particles in the star, given by an integer number $\bar m$ which can take positive and negative values. Static non-rotating solutions correspond to $\bar m=0$. In the rotating case, however, the parameter space to be considered is fairly small as free-field configurations with $\bar m\geq1$ can become dynamically unstable~\cite{PhysRevLett.123.221101,PhysRevD.102.101504,siemonsen2021stability,sanchis2021multifield,jaramillo2025full}. However, the wavelike nature of bosonic stars impacts the merger dynamics and associated GW emission due to the phase difference $\Delta\epsilon$ in the oscillation of the stars~\cite{Siemonsen:2023a,PhysRevD.106.124011}. This peculiar property results in an interference (constructive or destructive) that greatly changes the outcome of the merger. In addition, as shown in~\cite{Siemonsen:2023a}, the phase difference is necessary to form post-merger rotating scalar boson stars, in models with self-interacting fields where the rotating stars are dynamically robust.

In~\cite{PhysRevD.106.124011} we performed a systematic study of the impact of the relative phase on the fate of head-on collisions of rotating Proca stars. To do so, we performed hundreds of numerical-relativity simulations of both equal-  and unequal-mass configurations. Our results showed that understanding the effect of the phase is essential to correctly describe such systems. However, head-on collisions impose an unrealistic high degree of symmetry in the system. As a first step to relax this assumption, in this paper, we extend the set of simulations of~\cite{PhysRevD.106.124011} by adding orbital angular momentum in the initial data. We study the dynamics of about 100 eccentric mergers of equal-mass Proca star configurations with different masses, compactness, spins, phase, and initial boost. We shall assume small initial velocities to maintain the violations of the Hamiltonian and momentum constraints sufficiently low. Our simulations highlight the role played by the relative phase in the outcome of the mergers, leading to diverse scenarios. Those include the formation of black holes surrounded by transient Proca clouds, the formation of ``hypermassive'' Proca stars, or the appearance of dynamically-unstable spinning $\bar m$ = 2 Proca stars. For some models, odd parity GW modes are also triggered. Since those are absent in BBH mergers, they would offer a smoking-gun signature in potential GW searches of Proca stars.

This paper is organized as follows: Section~\ref{sec2} briefly describes the framework for our study, with a basic outline of the Einstein-(complex) Proca system. The procedure we follow to obtain initial data for the simulations is outlined in Section~\ref{sec3} as well as the specific numerical setup we employ. We report on our results in Section~\ref{results}. Finally, our conclusions are presented in Section~\ref{conclusions} along with some remarks on possible pathways for future research. Henceforth, units with $G=c=1$ are used.

%%%%%%%%%%%%%%%%%%%%%%
\section{Framework}
\label{sec2}
%%%%%%%%%%%%%%%%%%%%%%

We investigate the dynamics of Proca star binaries in eccentric orbits by solving numerically the Einstein-(complex, massive) Proca system, given by the action $\mathcal{S}=\int d^4x\sqrt{-g}\mathcal{L}$, where the Lagrangian density depends on the Proca potential $\mathcal{A}$ and the Proca field strength $\mathcal{F}=d\mathcal{A}$,
\begin{equation}
    \mathcal{L} = \frac{R}{16\pi}-\frac{1}{4}\mathcal{F}_{\alpha\beta}\bar{\mathcal{F}}^{\alpha\beta}-\frac{1}{2}\mu^2\mathcal{A}_{\alpha}\bar{A}^{\alpha}.
\end{equation}
In this equation the bar denotes complex conjugation, $R$ is the Ricci scalar and $\mu$ the Proca-field mass.
The metric in the $3+1$ decomposition is written in the following form
\begin{equation}
    ds^2=-\alpha^2dt^2+\gamma_{ij}\left(dx^i+\beta^idt\right)\left(dx^j+\beta^jdt\right)\,,
\end{equation}
where $\alpha$ is the lapse function, $\beta^i$ the shift vector, and $\gamma_{ij}$ is the three-dimensional spatial metric. We can define the operator projecting spacetime quantities onto the spatial hypersurfaces $\gamma^{\mu}_{\nu}=\delta ^{\mu}_{\nu}+n^{\mu}n_{\nu}$ (with $n^{\mu}$, a timelike unit vector orthogonal to the hypersurfaces). The Proca field is then split into the following $3+1$ quantities
\begin{equation}
    \begin{split}
         \mathcal{A}_{\mu}&=\mathcal{X}_{\mu}+n_{\mu}\mathcal{X}_{\phi}\,,\\
         \mathcal{X}_{i}&=\gamma^{\mu}_i\mathcal{A}_{\mu}\,,\\
         \mathcal{X}_{\phi}&=-n^{\mu}\mathcal{A}_{\mu}\,,
    \end{split}
\end{equation}
with $\mathcal{X}_{i}$ the vector potential and $\mathcal{X}_{\phi}$ the scalar potential.
Moreover, the three-dimensional ``electric" field $E^i$ and ``magnetic" field $B^i$ are introduced in analogy with Maxwell's theory,
\begin{equation}
E_i=\gamma^{\mu}_{\nu}\mathcal{F}_{\mu\nu}n^{\nu}\,,\quad\quad\quad B_i=\epsilon^{ijk}D_i\mathcal{X}_k \,,
\end{equation}
where $D_i$ is the covariant three-dimensional derivative with respect to $\gamma_{ij}$, $\epsilon^{ijk}$ is the three-dimensional Levi-Civita tensor, and $E_{\mu}n^{\mu}=B_{\mu}n^{\mu}=0$. The evolution equations (for further details see \cite{PhysRevD.106.124011}) are closed by the Hamiltonian and momentum constraint equations,
\begin{eqnarray}
       \mathcal{H}&=&R-K_{ij}K^{ij}+K^2-2\big(E^iE_i+B^iB_i \nonumber \\
&+&\mu^2\big(\mathcal{X}_{\phi}^2+\mathcal{X}^{i}\mathcal{X}_{i}\big)\big)=0\,,
\\
\mathcal{M}_i&=&D^jK_{ij}-D_iK-2\left(\epsilon^{ijk}E^jB^k+\mu^2\mathcal{X}_{\phi}\mathcal{X}_{i}\right)=0\,,
\end{eqnarray}
%\begin{equation}
%    \begin{split}
%        \mathcal{H}&=R-K_{ij}K^{ij}+K^2-2\big(E^iE_i+B^iB_i\\
%&+\mu^2\big(\mathcal{X}_{\phi}^2+\mathcal{X}^{i}\mathcal{X}_{i}\big)\big)=0\,,
%    \end{split}
%\end{equation}
%\begin{equation}
%    \begin{split}
%        \mathcal{M}_i&=D^jK_{ij}-D_iK-2\left(\epsilon^{ijk}E^jB^k+\mu^2\mathcal{X}_{\phi}\mathcal{X}_{i}\right)\,,
%    \end{split}
%\end{equation}
where $K_{ij}$ is the extrinsic curvature tensor and $K$ is its trace, $K=K^i_i$.

%%%%%%%%%%%%%%%%%%%%%%%%%%%%%%%%%%%%%%%%%%
%\subsection{Spinning Proca star solutions}
%%%%%%%%%%%%%%%%%%%%%%%%%%%%%%%%%%%%%%%%%%

Following~\cite{BRITO2016291} - see also~\cite{Herdeiro:2019mbz,Santos:2020pmh} -  we build spinning Proca star solutions with the following axisymmetric line element ansatz,
\begin{equation}
\begin{split}
    ds^2=&-e^{2F_0}dt^2+e^{2F_1}\left(dr^2+r^2d\theta^2\right)\\
    &+e^{2F_2}r^2\sin^2\theta\left(d\varphi-\frac{W}{r}dt\right)^2\,,
    \end{split}
\label{line-element}
\end{equation}
where $(r,\theta,\phi)$ are the spherical coordinates, $t$ is the time coordinate, and $F_0$, $F_1$, $F_2$ and $W$ are real functions of $\left(r,\theta\right)$. The Proca field $\mathcal{A}$ is given in terms of a different set of real functions of $\left(r,\theta\right)$,
%\begin{eqnarray}
%\mathcal{A}=\Big(\frac{H_1}{r}dr+H_2d\theta+iH_3\sin{\theta}d\varphi +
%    iVdt\Big)e^{i\left(\bar m\varphi-\omega t+\epsilon\right)}
%\end{eqnarray}
\begin{equation}
\begin{split}
    \mathcal{A}=&\Big(\frac{H_1}{r}dr+H_2d\theta+iH_3\sin{\theta}d\varphi +\\
    &iVdt\Big)e^{i\left(\bar m\varphi-\omega t+\epsilon\right)}
\end{split}
\label{eq:field_ansatz}
\end{equation}
with $\bar m\in \mathbb{Z}$, $\omega$ is the field frequency and $\epsilon$ the initial phase of the star. For isolated configurations, the dependence on the internal phase structure of the star $\bar m\varphi-\omega t+\epsilon$ disappears at the level of the stress-energy tensor, which depends solely on $(r,\theta)$ through $\mathcal{A}_{\alpha}\mathcal{\bar A}^{\alpha}$ and other quadratic contributions. Therefore, the solution is axisymmetric and stationary (see Eq.~(\ref{line-element})). However, when a second star is present, the difference between their internal phase structures, given by $|\bar m_1\varphi-\omega_1 t+\epsilon_1 - \bar m_2\varphi-\omega_2 t+\epsilon_2|$, plays a pivotal role. In equal-mass collisions, this difference only depends on the phases $\epsilon_1,\epsilon_2$ and remains constant from the initial configuration to merger, and is defined as $\Delta \epsilon = |\epsilon_1 - \epsilon_2|$. This parameter is crucial for classifying the binaries and determining their properties, as it alters the way the stars interact, leading to interference phenomena reminiscent of wave-like behavior. Thus, while the stars and their individual stress-energy tensors are axisymmetric, the fields exhibit an interaction driven by their internal phase structures. 

We refer to Figure $1$ of \cite{PhysRevD.106.124011} to depict the domain of existence and the compactness of the spinning $\bar m=1$, nodeless Proca star solutions. The frequency of stable solutions is restricted to the interval between $\omega/\mu=1$ to $\omega/\mu\sim0.711$, as below this last value Proca star solutions develop a light ring pair \cite{PhysRevLett.119.251102} creating a spacetime instability \cite{PhysRevLett.130.061401}. We note that the particle mass $\mu$ plays the role of a fundamental scale of the system and each physical quantity can be appropriately rescaled by $\mu$.  Since bosonic stars do not have a surface discontinuity of the energy density (contrary to fermionic stars), their compactness is usually defined as the ratio of the radius $R_{99}$ that contains $99\%$ of the mass $M_{99}$, 
$C=2M_{99}/R_{99}$.
%The bottom panel of Figure $1$ in \cite{PhysRevD.106.124011} shows the compactness to be inverse proportionally to the frequency $\omega$. 

%The transformation between the function introduced in Eqs.~(\ref{line-element}) and (\ref{eq:field_ansatz}) and the $3+1$ Proca fields variables is given as follows:
%\begin{eqnarray}
%    \alpha&=&e^{F_0}\,\quad\quad \beta^{\varphi}=\frac{W}{r} \\       \gamma_{rr}&=&e^{2F_1},\, \gamma_{\theta\theta}=e^{2F_1}r^2,\, \gamma_{\varphi\varphi}=e^{2F_2}r^2\sin^2{\theta}\\
%    \chi_{\phi}&=&-n^{\mu} \mathcal{A}_{\mu}\,\quad\quad \chi_{i}=\gamma^{\mu}_i\mathcal{A}_{\mu} \\
%    E^i\nonumber&=&-i\frac{\gamma^{ij}}{\alpha}\left[D_j\left(\alpha\chi_{\phi}\right)+\partial_t\chi_j\right]
%\end{eqnarray}

%%%%%%%%%%%%%%%%%%%%%%%%%%%%%%%%%%%
\section{Initial data and numerics}
\label{sec3}
%%%%%%%%%%%%%%%%%%%%%%%%%%%%%%%%%%%

As in~\cite{PhysRevD.106.124011} initial data for binary Proca stars is built through a superposition of two rotating stars described by the same Proca field (implying a common value for the boson mass $\mu$), such that:
 \begin{eqnarray}
        \mathcal{A}(x_i)&=&\mathcal{A}^{(1)}(x_i-x_0)+\mathcal{A}^{(2)}(x_i+x_0)\,,\\
        \gamma_{ij}(x_i)&=&\gamma_{ij}^{(1)}(x_i-x_0)+\gamma_{ij}^{(2)}(x_i+x_0) \nonumber \\
        && -\gamma_{ij}^{\text{PS}}(\pm x_0)\,,\\
        \alpha(x_i)&=&\alpha^{(1)}(x_i-x_0)+\alpha^{(2)}(x_i+x_0) \nonumber \\
        && -1\,.
\end{eqnarray}
In these equations we label the stars with superscripts $(1)$ and $(2)$ and indicate their respective initial position with $\pm x_0$. The initial separation of the two stars is given by the coordinate distance $D\mu=50$ (in particular $\pm x_0\mu=25$). Moreover,$\gamma_{ij}^{\text{PS}}(x_0)$ is the 3-metric evaluated at the center of a given star with $\gamma_{ij}^{\text{PS}}(x_0)=\gamma_{ij}^{(1)}(+x_0)=\gamma_{ij}^{(2)}(-x_0)$.

Stationary fundamental bosonic stars are described by the oscillation frequency $\omega/\mu$ of the field, which determines the dimensionless mass $M\mu$ and angular momentum $J\mu^2$. These physical quantities are computed through the Komar integrals,
\begin{eqnarray}
    M&=&-\int_{\Sigma}drd\theta d\varphi\left(2T^t_t-T^{\mu}_{\mu}\right)\alpha\sqrt{\gamma},\label{eq:Komarmass}\\
    J&=&\int_{\Sigma}drd\theta d\varphi \,T^{t}_{\varphi}\,\alpha\sqrt{\gamma}\label{eq:Komarmom},
\end{eqnarray}
where $T_{\mu\nu}$ is the stress energy tensor of the Proca field (we refer to \cite{PhysRevD.106.124011} for its explicit definition) and $\gamma$ is the determinant of the 3-metric. 
We note that the angular momentum of bosonic stars is quantized. It is related to the  Noether charge $Q$ of the star, counting the number of bosonic particles, by the relation $J=\bar mQ$. An infinitesimal loss  or gain of angular momentum corresponds to a loss or gain of particles. A compelling implication of this quantization is that if a star loses angular momentum faster than it loses mass, it will eventually be stripped of all its angular momentum~\cite{PhysRevLett.123.221101,di2020dynamical}.

Each star is characterized by its oscillation frequency, $\omega_1/\mu$ for the first star and $\omega_2/\mu$ for the second. Our dataset consists of equal-mass Proca star binaries (corresponding to $\omega_1=\omega_2=\omega$) with three distinct frequency values, $\omega/\mu=\lbrace0.83$, $0.85$,  $0.87\rbrace$, and constant initial coordinate separation, $D\mu=50$. To obtain eccentric orbits on the $xy$-plane, each star is given an initial boost along the $y$-axis. We explore a range of boost velocities, $v/c=\lbrace0.015, 0.02, 0.03, 0.04, 0.05\rbrace$, and vary the relative phase $\Delta\epsilon$ from 0 to $\pi$ in increments of $\pi/6$ (i.e.~$\Delta\epsilon=n\pi/6$ with $n=0,\dots,6$). In total, we have performed about 100 eccentric merger simulations to assess the impact of the initial boost and phase on the merger dynamics and GW emission. Table \ref{tab:simulations} reports the specific subset of initial data corresponding to the simulations that are explicitly discussed in this work.

\begin{table*}[t]
\caption{Specific models of the eccentric, equal-mass, Proca star binaries employed in the simulations discussed in this paper. From left to right the columns report: the name of the model,  the compactness of each star ($C$), the frequency of the star ($\omega/\mu$), the initial boost ($v/c$), the relative phase difference ($\Delta\epsilon$), and the phase of each individual star ($\epsilon_1$ and $\epsilon_2$).}
\label{tab:simulations}
\centering
\begin{tabular}{ccccccccc}
%\toprule 
\hline
Model &$M\mu$&$J\mu^2$& $C$ & $\omega/\mu$ & $v/c$ & $\Delta\epsilon$ & $\epsilon_1$ & $\epsilon_2$\\
\hline 
$83$A$0$a&0.8940& 0.9440 & 0.2045 & $0.8300$ & $0.015$ & $0$ & $0$ & $0$ \\
$83$A$0$b&0.8940& 0.9440 &  0.2045& $0.8300$ & $0.015$ & $0$ & $\pi/4$ & $\pi/4$ \\
$83$A$0$c&0.8940& 0.9440 & 0.2045 & $0.8300$ & $0.015$ & $0$ & $\pi/2$ & $\pi/2$ \\
$83$B$0$&0.8940& 0.9440 & 0.2045 & $0.8300$ & $0.020$ & $0$ & $0$ & $0$ \\
$83$B$6$&0.8940& 0.9440 & 0.2045 & $0.8300$ & $0.020$ & $\pi$ & $0$ & $\pi$ \\
$83$C$0$&0.8940& 0.9440 & 0.2045 & $0.8300$ & $0.030$ & $0$ & $0$ & $0$ \\
$83$C$6$&0.8940& 0.9440 & 0.2045 & $0.8300$ & $0.030$ & $\pi$ & $0$ & $\pi$ \\
\hline 
$85$A$0$&0.8536& 0.8960 & 0.1806 & $0.8500$ & $0.015$ & $\pi$ & $0$ & $\pi$ \\
$85$A$6$&0.8536& 0.8960  & 0.1806 & $0.8500$ & $0.015$ & $\pi$ & $0$ & $\pi$ \\
$85$C$0$&0.8536& 0.8960  & 0.1806 & $0.8500$ & $0.030$ & $0$ & $0$ & $0$ \\
$85$C$3$&0.8536& 0.8960  & 0.1806 & $0.8500$ & $0.030$ & $\pi/2$ & $0$ & $\pi/2$ \\
$85$C$6$&0.8536& 0.8960  & 0.1806 & $0.8500$ & $0.015$ & $\pi$ & $0$ & $\pi$ \\
$85$A$9$&0.8536& 0.8960  & 0.1806 & $0.8500$ & $0.015$ & $3\pi/2$ & $0$ & $3\pi/2$ \\
\hline
$87$A$0$&0.8077& 0.8426 & 0.1566 & $0.8700$ & $0.015$ & $0$ & $0$ & $0$ \\
$87$B$0$&0.8077& 0.8426 & 0.1566 & $0.8700$ & $0.020$ & $0$ & $0$ & $0$ \\
$87$B$2$&0.8077& 0.8426 & 0.1566 & $0.8700$ & $0.020$ & $\pi/3$ & $0$ & $\pi/3$ \\
$87$B$5$&0.8077& 0.8426 & 0.1566 & $0.8700$ & $0.020$ & $5\pi/6$ & $0$ & $5\pi/6$ \\
$87$B$6$&0.8077& 0.8426 & 0.1566 & $0.8700$ & $0.020$ & $\pi$ & $0$ & $\pi$ \\
$87$C$0$&0.8077& 0.8426 & 0.1566 & $0.8700$ & $0.030$ & $0$ & $0$ & $0$ \\
$87$C$2$&0.8077& 0.8426 & 0.1566 & $0.8700$ & $0.030$ & $\pi/3$ & $0$ & $\pi/3$ \\
$87$C$3$&0.8077& 0.8426 & 0.1566 & $0.8700$ & $0.030$ & $\pi/2$ & $0$ & $\pi/2$ \\
$87$C$4$&0.8077& 0.8426 & 0.1566 & $0.8700$ & $0.030$ & $2\pi/3$ & $0$ & $2\pi/3$ \\
$87$C$5$&0.8077& 0.8426 & 0.1566 & $0.8700$ & $0.030$ & $5\pi/6$ & $0$ & $5\pi/6$ \\
$87$C$6$&0.8077& 0.8426 & 0.1566 & $0.8700$ & $0.030$ & $\pi$ & $0$ & $\pi$ \\
$87$D$6$&0.8077& 0.8426 & 0.1566 & $0.8700$ & $0.040$ & $\pi$ & $0$ & $\pi$ \\
$87$E$6$&0.8077& 0.8426 & 0.1566 & $0.8700$ & $0.050$ & $\pi$ & $0$ & $\pi$ \\
\hline
%\bottomrule
\end{tabular}
\end{table*} 

%%%%%%%%%%%%%%%%%%
%\section{Numerics}
%%%%%%%%%%%%%%%%%%

Our simulations have been performed with the open-source code \textsc{einstein toolkit} \cite{ET_web, Löffler_2012} with the \textsc{cactus} framework and using mesh refinement. The time-dependent differential equations are integrated with the method of lines employing a fourth-order Runge-Kutta scheme. The left-hand side of the Einstein equations is solved using the \textsc{lean} code, based on the strongly-hyperbolic $3+1$ BSSN formulation. The Proca equations, as described in \cite{PhysRevD.106.124011}, are solved using the code of \cite{Zilhão_2015, CODE_PS_web, witek_2023_7791842} and extended to take into account a complex field in~\cite{PhysRevLett.123.221101, PhysRevD.99.024017}. The simulations assume equatorial plane symmetry in a grid with seven refinement levels. The spatial (coordinate) domain of each level is $\{320.0, 96.0, 96.0, 48.0, 48.0, 6.0, 2.0\}/\mu$, with a corresponding spatial resolution for each level of $\{4.0, 2.0, 1.0, 0.5, 0.25, 0.125, 0.0625\}/\mu$. 

In order to extract the outgoing gravitational radiation produced in the mergers we adopt the Newman-Penrose (NP) spin coefficients formalism \cite{1962JMP.....3..566N},  
%in the limit of radiative empty space-time, for large value of a radial coordinate $r$, the Riemann tensor exhibits a characteristic asymptotic behaviour, being the series expansion in negative powers of $r$, namely $\Psi_n=\mathcal{O}\left(r^{n-5}\right)$, breaking the space-time in five regions, such that in the radiative zone only the $\Psi_4$ term remains dominant and the Riemann tensor is eseentially null. This is the so-called Newman-Penrose (NP) scalar and it is computed by the Weyl tensor  $C_{\alpha\beta\gamma\delta}$:
with the $\Psi_4$ NP scalar computed from the Weyl tensor  $C_{\alpha\beta\gamma\delta}$.  %$\Psi_4=C_{\alpha\beta\gamma\delta}n^{\alpha}\overline{m}^{\beta}n^{\gamma}\overline{m}^{\delta}$.
%
%where $n^{\mu}$ and $\overline{m}^{\mu}$ are part of the null tetrad of vectors $z_{m_{\mu}}=\left(k_{\mu}, n_{\mu}, m_{\mu}, \overline{m}_{\mu}\right)$. The real null vectors $k_{\mu}$ and $n_{\mu}$ are the outgoing and ingoing null normal vector to the extraction worldtube $\Gamma$ of the GWs radiation, while the vector $m_{\mu}$ and its complex conjugate $\overline{m}_{\mu}$ are tangent to $\Gamma$ at constant time slicing and defined from a pair of real, orthogonal unit spacelike vectors, which in spherical coordinates, as standard choice, are the coordinate basis vectors $e^{\mu}_{\left(\theta\right)}$ and $e^{\mu}_{\left(\varphi\right)}$ and such that: $m^{\mu}=\frac{1}{\sqrt{2}}\left(e^{\mu}_{\left(\theta\right)} + i e^{\mu}_{\left(\varphi\right)}\right)$. These vectors satisfy the following orthonormal properties: $k_{\mu}n^{\mu}=-1$ and $m_{\mu}\overline{m}^{\mu}=1$, while the other inner products are zero.
%
At a given extraction radius $r_{\rm ext}$ it is useful to perform a decomposition of the NP scalar onto a $-2$ spin-weighted spherical harmonics basis,
\begin{equation}
    \Psi_4\left(t,r_{\rm ext},\theta,\varphi\right)=\sum_{l,m}\psi^{l,m}_4\left(t,r_{\rm ext}\right) \text{ }^{-2}Y_{l,m}\left(\theta, \varphi\right)\,.
\end{equation}
As discussed below, our numerical simulations display the dominance of the  quadrupolar modes $\left(l,m\right)=\left(2,\pm2\right)$  followed by the weaker $\left(2,0\right)$, $\left(3,\pm3\right)$ and $\left(3,\pm2\right)$ modes.

\begin{figure}[t!]
  \centering
      \includegraphics[width=1.0\linewidth]{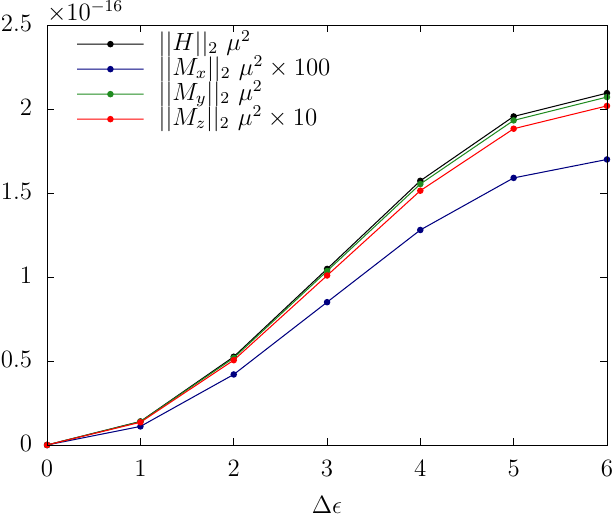}
      \makeatletter\long\def\@ifdim#1#2#3{#2}\makeatother
\caption{Difference with respect to the zero-dephase configuration of the $L_2$-norm of the Hamiltonian constraint (black) and of the momentum constraint ($x$ component in blue, $y$ component in green and $z$ component in red) for equal-mass Proca star mergers with $\omega/\mu=0.83$ and initial boost $v/c=0.03$, as a function of the relative phase $\Delta \epsilon$ (in units of ${\pi}/{6}$).}
\label{fig:Con_vs_epsilon_omega08300_v030}
\end{figure}

Since Einstein's equations are non-linear, the construction of binary initial data from a superposition of two compact stars leads to some violation of the Hamiltonian and momentum constraints. Figure \ref{fig:Con_vs_epsilon_omega08300_v030} shows the $L_2$-norm of these two constraints at initial time for an  initial boost $v/c=0.03$ and for the more compact set of stars of our sample ($\omega/\mu=0.83$). The $L_2$-norm is plotted as a function of the relative phase $\Delta \epsilon$ in units of $\pi/6$. In the plot we display the difference on the constraint violations with respect to the zero-dephase configuration. The violations increase with the phase difference with roughly the same slope.
%(configuration with the maximum violations approximately of $7\times10^{-5}$ for Hamiltonian and $10^{-6}$, $10^{-6}$ and $6\times10^{-7}$ for respectively the $x$, $y$ and $z$ components of the Momentum constraint)

\begin{figure}[t]
  \centering
      \includegraphics[width=1.0\linewidth]{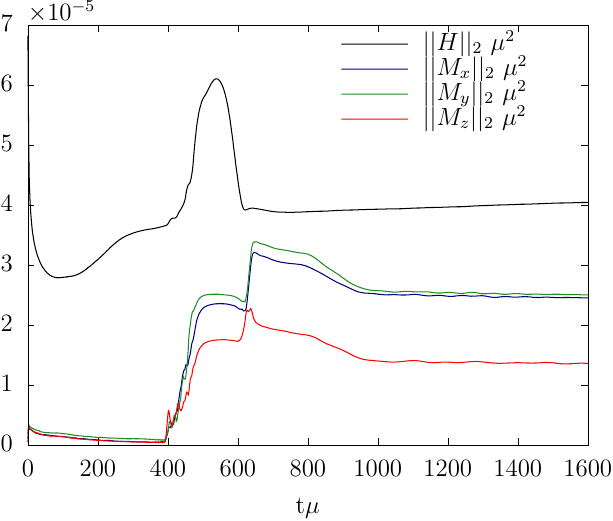}
      \makeatletter\long\def\@ifdim#1#2#3{#2}\makeatother
\caption{Evolution of the $L_2$-norm of the Hamiltonian constraint (black) and of the momentum constraint ($x$ components in blue, $y$ components in green and $z$ components in red) for an equal-mass Proca star merger with $\omega/\mu=0.87$ and initial boost $v/c=0.03$. After the maximum violations are reached at merger, the constraints flatten to smaller asymptotic values of a few times $10^{-5}$.}
\label{fig:Con_vs_time_diff00_omega08700_v030}
\end{figure}

Figure \ref{fig:Con_vs_time_diff00_omega08700_v030} gives an illustrative example the evolution of the $L_2$-norms for both the Hamiltonian and momentum constraints during an equal-mass Proca star merger. In this simulation, the stars have a frequency of $\omega/\mu=0.87$, no relative phase difference, and an initial boost of $v/c=0.030$. The largest constraint violations occur during the late inspiral and merger phases, after which they settle to almost constant values of approximately $4\times10^{-5}$ for the Hamiltonian constraint and about $2.5\times10^{-5}$, $2.5\times10^{-5}$, and $1.5\times10^{-5}$ for the $x$, $y$, and $z$ components of the momentum constraint, respectively.

\begin{figure*}[t]
\subfloat{
  \includegraphics[width=0.50\linewidth]{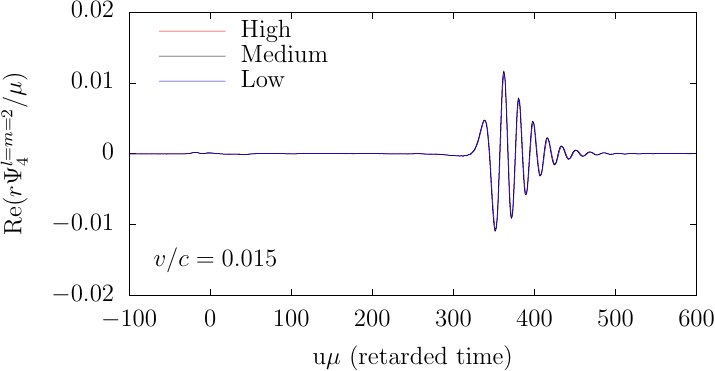}
}
\subfloat{
  \includegraphics[width=0.50\linewidth]{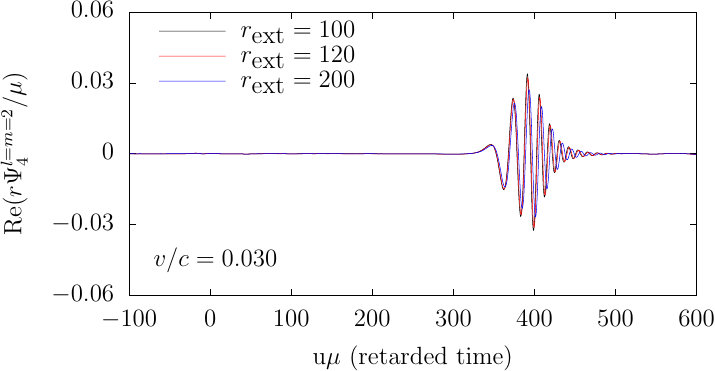}
}\\
%\vspace{-2.5cm}
\subfloat{
  %\hspace{-0.58cm}\includegraphics[width=0.54\linewidth]{Plot/GW_e00_v015_comparedx_residual_r100.pdf}
  \hspace{-0.58cm}\includegraphics[width=0.54\linewidth]{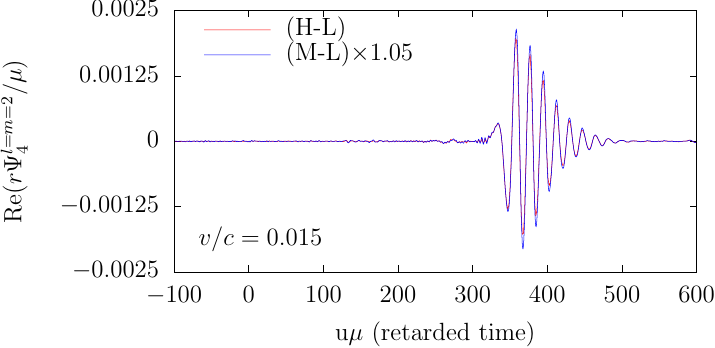}
}
\subfloat{
  %\hspace{-0.3cm}\includegraphics[width=0.50\linewidth]{Plot/GW_e00_v030_comparedx_residual_r100.pdf}
  \hspace{-0.3cm}\includegraphics[width=0.50\linewidth]{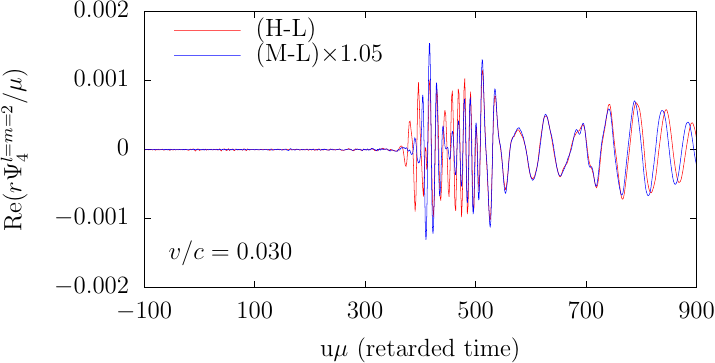}
}
\caption{Top-left panel: $r\Psi_4^{22}$ mode for simulation $85$A$0$ at extraction radius $r_{\rm ext}=100$. Top-right panel: $r\Psi_4^{22}$ mode for simulation $85$C$0$ at several extraction radii ($r_{\rm ext}=100$, $120$ and $200$). Bottom row: convergence study for simulation $85$A$0$ (left) and $85$C$0$ (right). The waveforms are artificially rescaled according to fourth-order convergence.}
\label{fig:GW_e00_v015_rextdx}
\end{figure*}

We turn now to discuss the accuracy of our numerical simulations and the robustness of our results. The interested reader is addressed to \cite{ Zilhão_2015,PhysRevLett.123.221101, PhysRevD.99.024017} for a convergence analysis of the \textsc{einstein toolkit} evolving Proca fields, Proca stars and Proca binaries. We start analyzing the role of the extraction radius and the spatial resolution. The top-left panel of Figure \ref{fig:GW_e00_v015_rextdx} plots the $l=m=2$ mode for model $85$A$0$ at the extraction radius $r_{\rm ext}\mu=100$ with three different resolutions: $3.0$ (high), $4.0$ (medium) and $8.0$ (low) with respect to the retarded time $\rm u\mu=\left(t-r_{\rm ext}\right)\mu$. At this extraction radius all three waveforms overlap regardless of the resolution employed. A small amount of junk radiation is present (yet hardly visible) before the physical waveform begins. In all cases this junk radiation happens sufficiently early on such that it can be easily separated from the actual physical emission which remains unaffected by this artifact of the initial data. On the top-right panel of the same figure we compare three waveforms extracted at different radii ($r_{\rm ext}\mu=100$, $120$ and $200$) for model $85$C$0$. The evolution of this model is done using the highest resolution (since the outcome is significantly more sensitive to the numerical resolution). As we can see the extraction of the waveform is more accurate the further this radius is from the source. As a result, we select $r_{\rm ext}\mu=100$ as our fiducial choice of extraction radius. The bottom row of Figure~\ref{fig:GW_e00_v015_rextdx} shows a convergence study for the two models ($85$A$0$ on the left and $85$C$0$ on the right). The waveforms have been conveniently shifted and rescaled for an analytical fourth-order convergence. The comparison shows that the order of convergence of our numerical results lies between three and four.

%%%%%%%%%%%%%%%%%%%%%
\section{Results}
\label{results}
%%%%%%%%%%%%%%%%%%%%%

\begin{figure*}[t]
  \centering
  \subfloat{\includegraphics[width=0.5\linewidth]{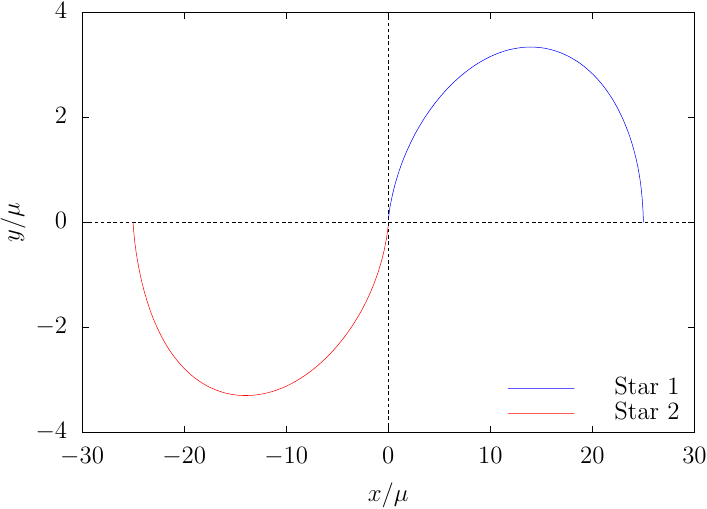}}
  \subfloat{\includegraphics[width=0.5\linewidth]{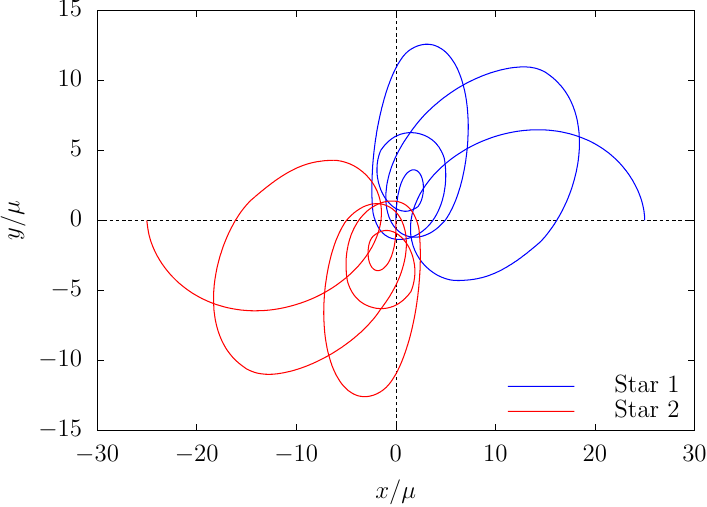}}
\caption{Trajectories of PSs in binaries for simulations $83$A$0$a (left) and $87$C$6$ (right) in the equatorial plane $xy$.}
\label{fig:trajectories_BBS}
\end{figure*}

As mentioned before, we have performed around $100$ simulations of eccentric inspirals of equal-mass spinning Proca stars, with different initial boosts and relative phases. This large sample of simulations allows us to explore in a systematic way the dynamics of such systems and the GW signals emitted. The variety of possible post-merger remnants and waveforms might contribute to ongoing efforts to analyze LVK events from the perspective of mergers of exotic compact objects~\cite{PhysRevLett.126.081101,Bustillo:2023,evstafyeva2024gravitational}. 
The simulations performed in this work have focused on dynamically robust $\bar m=1$ Proca stars with $\left(\omega/\mu,M\mu,J\mu^2\right)=\left(0.8300, 0.8940, 0.9440 \right)$, $\left(0.8500, 0.8536, 0.8960 \right)$ and $\left(0.8700, 0.8077, 0.8426 \right)$. The two Proca stars merge either in less than one complete orbit or after more than one encounter, depending on the internal phase structure of the Proca stars (namely, the relative phase) and the values of the initial boost. As illustrative examples Figure~\ref{fig:trajectories_BBS} displays the trajectories of each star in the equatorial plane for models $83$A$0$a (grazing collision) and $87$C$6$ (multiple encounter). As we discuss below, for some of the models the remnant object acquires a non-negligible kick, due to the non-zero relative phase. Moreover, for most models the end state is a Kerr black hole surrounded by a Proca field remnant (a Proca cloud) storing a fraction of the initial Proca mass and angular momentum. The mass of this remnant can be initially comparable to that of the black hole, which has an important imprint in the emission of GWs. In a few other cases, the outcome of the merger is an unstable hypermassive rotating Proca star, which gradually loses angular momentum and collapses onto a black hole. The imprint this process leaves on the emitted GW differs, however, from the burst-like signals observed for most models.

%%%%%%%%%%%%%%%%%%%%%%%%%%%%%%%%%%%%%%%%%%%%%%
\subsection{The role of the orbital angular momentum}
\label{roleangularmomentum}
%%%%%%%%%%%%%%%%%%%%%%%%%%%%%%%%%%%%%%%%%%%%%%
%\NS{Change everywhere initial AM by  orbital AM}
\begin{figure*}
  \centering
   \subfloat{
  \includegraphics[width=0.5\linewidth]{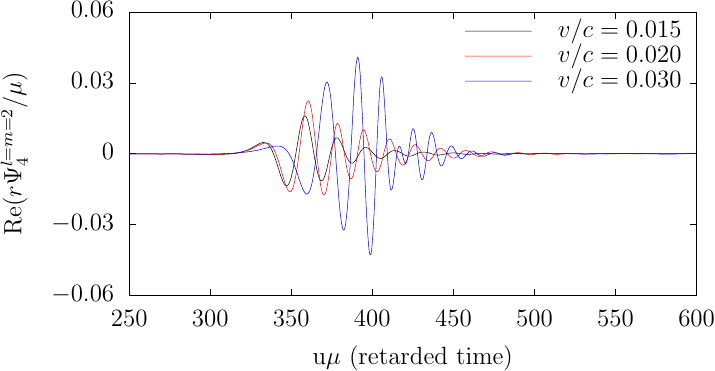}
}
\subfloat{
  \centering
  \includegraphics[width=0.5\linewidth]{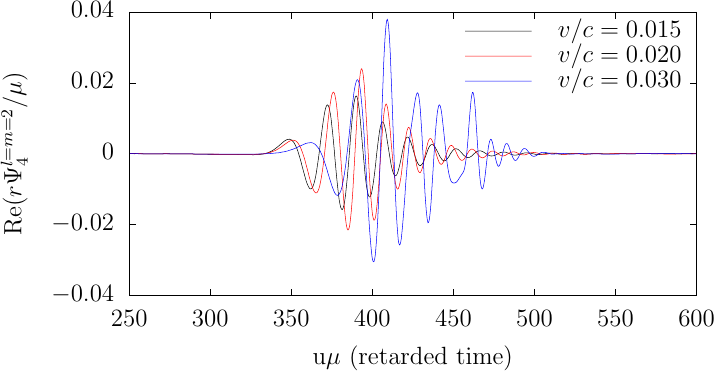}
}

\caption{$r\Psi_4^{22}$ mode for in-phase equal-mass mergers (left $\omega/\mu=0.83$; right $\omega/\mu=0.87$), varying the initial boost ${v}/c=0.015$ (black), $0.02$ (red) and $0.03$ (blue).} 
\label{fig:GW_comparev_e00}
\end{figure*}

\begin{figure}
  \centering
  \includegraphics[width=1\linewidth]{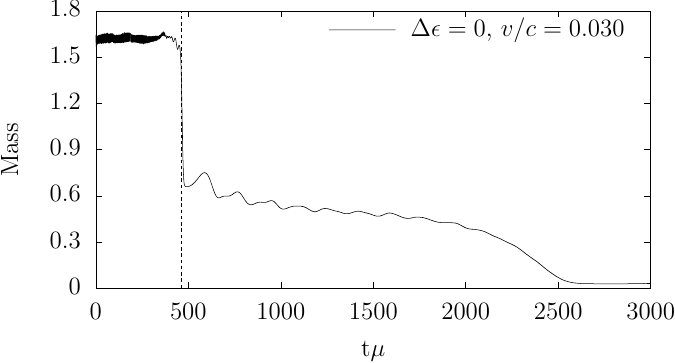}\\
  \includegraphics[width=1\linewidth]{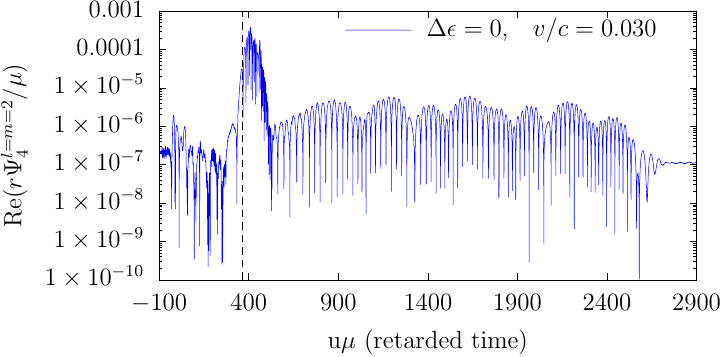}
\caption{Top panel: Evolution of the integrated mass on the full radial extend of the grid for model $87$C$0$. Bottom panel: $r\Psi_4^{22}$ mode for the same model. The vertical dashed black line indicates the merger time ($\rm u\mu\sim360$).}
\label{fig:Mass_e00_v030_omega087}
\end{figure}

\begin{figure}[t]
  \centering
   \subfloat{
  \includegraphics[width=1\linewidth]{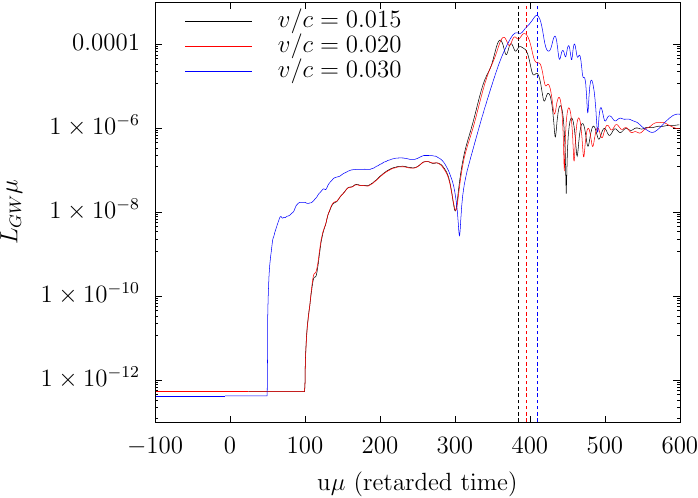}
} \\
\subfloat{
  \includegraphics[width=1\linewidth]{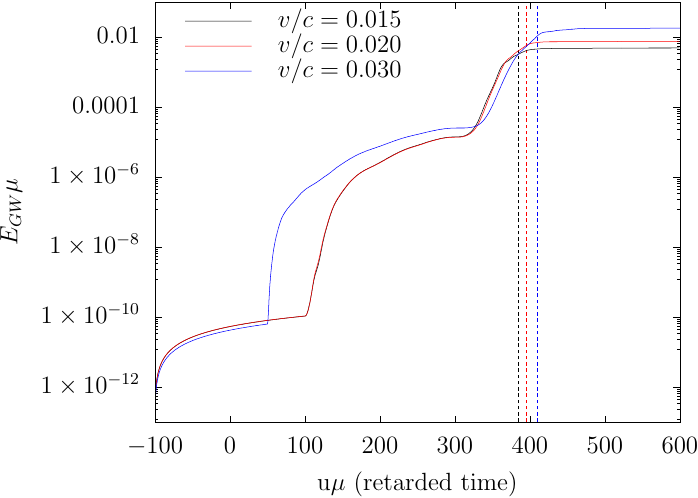}
}
\caption{Evolution of the GW luminosity (top panel) and integrated total emitted energy (bottom panel) for in-phase equal-mass Proca star mergers with $\omega/\mu=0.8700$ with initial boosts $v/c=0.015$  (model $87$A$0$, black), $0.02$  (model $87$B$0$, red) and $0.03$ (model $87$C$0$, blue). The vertical dashed lines indicate the time of merger: $\rm u\mu\sim 385$ for model $87$A$0$  (black), $\rm u\mu\sim395$ model $87$B$0$ (red) $\rm u\mu\sim410$, and model $87$C$0$ (blue).}
\label{fig:EGW_LGW_e00_v030_omega087}
\end{figure}

Adding an initial boost to the two stars adds non-zero orbital angular momentum in the system. In order to isolate the effects of the boost on the merger dynamics and GW emission we first consider only in-phase equal-mass systems ($\omega/\mu=0.83$ and $0.87$) separated by $D\mu=50$, and the three values of the boost,  $v/c=0.015$, $0.020$ and $0.030$. In Fig.~\ref{fig:GW_comparev_e00} we compare the $l=m=2$ mode of the NP scalar $\Psi_4$ for this set of models. The figure illustrates how the initial orbital angular momentum affects not only the morphology of the GW (increasing its amplitude for larger values of the boost) but also the duration of its post-merger phase, due to the additional momentum that delays the gravitational collapse. The final result of the merger does not depend on the initial boost, at least for our set of values. For smaller values of the boost, the merger takes less than a full orbit and it is followed by the prompt formation of a Kerr black hole. For a sufficiently large initial boost ($v/c=0.03$, models $83$C$0$ and $87$C$0$) a longer post-merger signal is visible. In such situations, the black hole that is formed right after merger is surrounded by a bosonic cloud which is dragged around by the angular momentum of the system. This corresponds to a short-lived Proca field remnant that, after losing part of its angular momentum, falls into the black hole.
 
This is clearly shown in Figure \ref{fig:Mass_e00_v030_omega087}. The top panel exhibits the evolution of the Proca mass, integrated on the total radial extension of the grid, for model $87$C$0$, using the Komar mass expression from Eq.~(\ref{eq:Komarmass}). After merger, which is indicated by a vertical dashed black line and almost in coincidence with black hole formation, there is a large sudden drop of the total mass. Due to the system’s significant angular momentum and because spinning Proca stars exceed the Kerr limit, not all of the Proca field collapses into the black hole. Instead, a transient Proca cloud forms around it, as evidenced by the fact that the total mass of the initial Proca star binary does not drop to zero. The black hole gradually absorbs the remaining Proca field, undergoing a transient phase that culminates in a ringdown. This sequence is imprinted on the gravitational-wave signal, as shown on a logarithmic scale in the bottom panel of Fig.~\ref{fig:Mass_e00_v030_omega087}. 

Previous studies such as~\cite{PhysRevD.77.044036} have reported similar dynamics for orbital mergers of scalar boson stars. Figure 2 in \cite{PhysRevD.77.044036} summarizes the diverse outcomes observed in these mergers. At sufficiently low initial boost velocities, the merger results in the formation of a Kerr black hole; however, at higher boost values, the system produces a transient, rotating bar of scalar field. This structure can either shed angular momentum and settle into a non-spinning boson star or split into two identical objects that eventually become unbound.
%Since the BS angular momentum is quantised, after the merger the remnant angular momentum is not enough to settle a stable, rotating and stationary BS. 
Almost all angular momentum is lost through GW emission, while the mass is roughly unchanged. The higher the increase in initial velocity, the smaller the ejecta. Further analysis of the dynamics of boson star mergers was reported in~\cite{Bezares_2018, PhysRevLett.123.221101}.

The dependence of the GW emission with the orbital angular momentum becomes more apparent when studying the GW luminosity, given by the expression:
\begin{equation}\label{eq:LGW}
    L_{\rm{GW}}=\frac{dE}{dt}=\lim_{r\to\infty}\frac{r^2}{16\pi}\sum_{l=2}^{\infty}\sum_{m=-l}^l\bigg|\int_{-\infty}^tdt'\Psi_4^{lm}\bigg|^2\,.
\end{equation}
In Figure \ref{fig:EGW_LGW_e00_v030_omega087} we plot the time evolution of the GW luminosity and emitted energy for an in-phase equal-mass Proca star merger with $\omega/\mu=0.8700$ for several values of the initial boost. Larger values of initial orbital angular momentum result in a more luminous event with more energy emitted through GW.

\begin{figure}[t]
  \centering
  \includegraphics[width=1.0\linewidth]{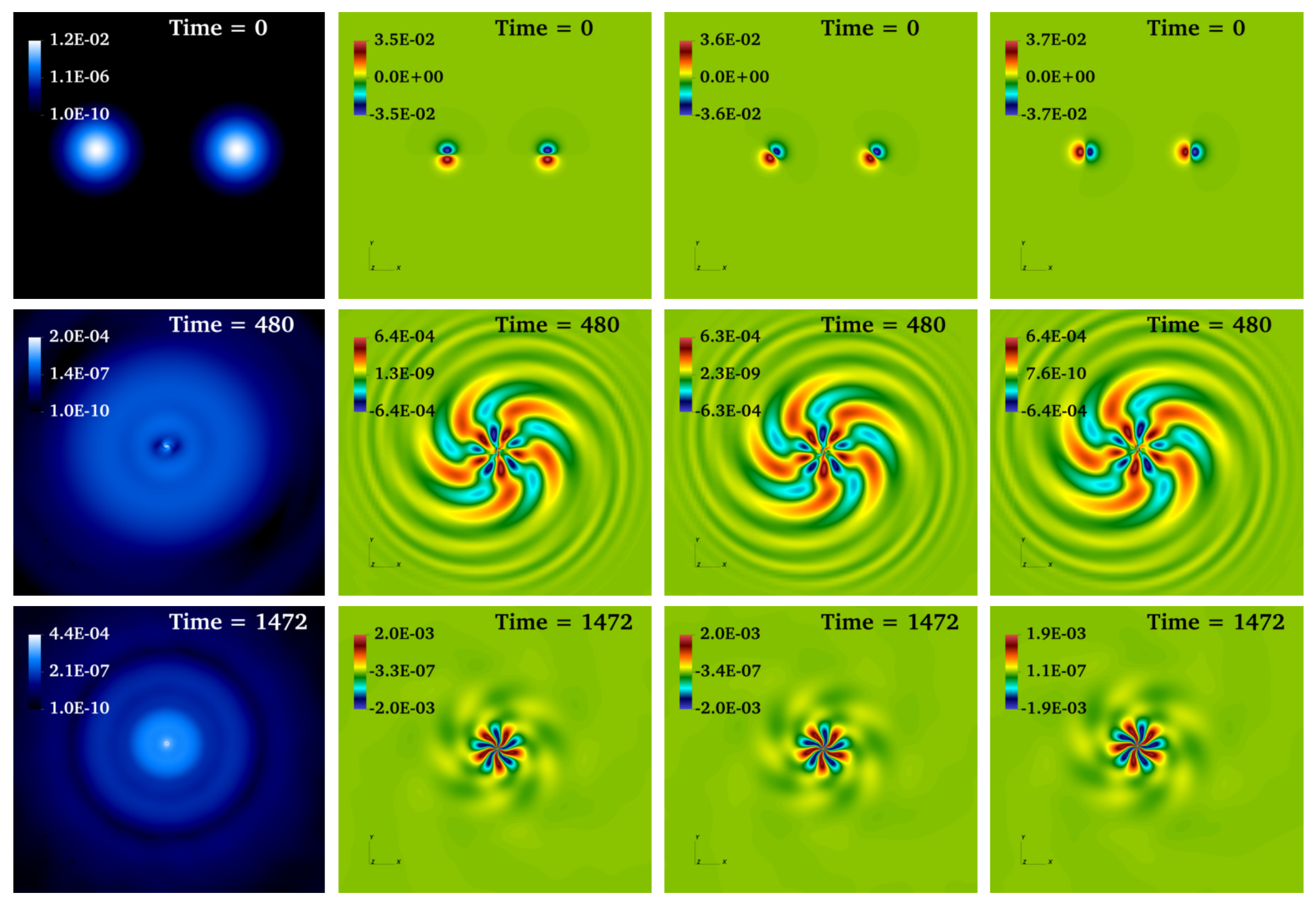}
\caption{Left column: equatorial plane snapshots of the energy density for model $83$A$0$a. Remaining columns: real part of scalar potential $\chi_{\phi}$ varying the phase of both stars (models $83$A$0$a, $83$A$0$b and $83$A$0$c, from left to right).}
\label{fig:chiphi_ep}
\end{figure}

\begin{figure}[t]
  \centering
    \includegraphics[width=0.5\textwidth]{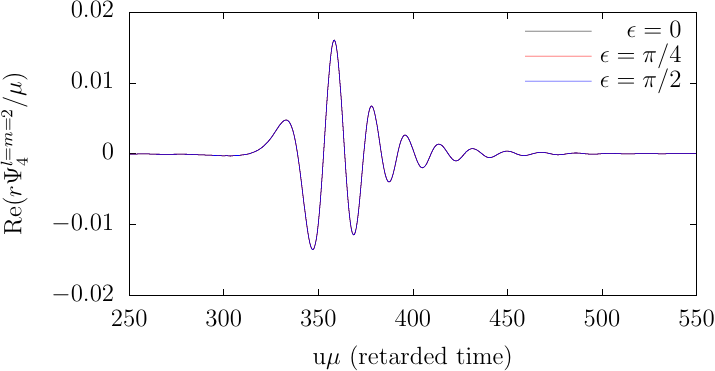}
\caption{Evolution of $r\Psi_4^{22}$ for models $83$A$0$a (black), $83$A$0$b (red) and $83$A$0$c (blue) with zero relative phase, showing that the waveforms perfectly overlap.}
\label{fig:e00_v015_ep}
\end{figure}

\begin{figure}[t]
\subfloat{
  \centering
  \includegraphics[width=1.0\linewidth]{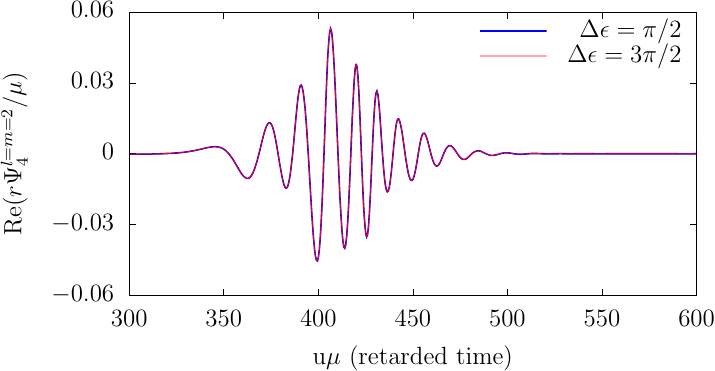}
}\\
\subfloat{
  \centering
  \includegraphics[width=1.0\linewidth]{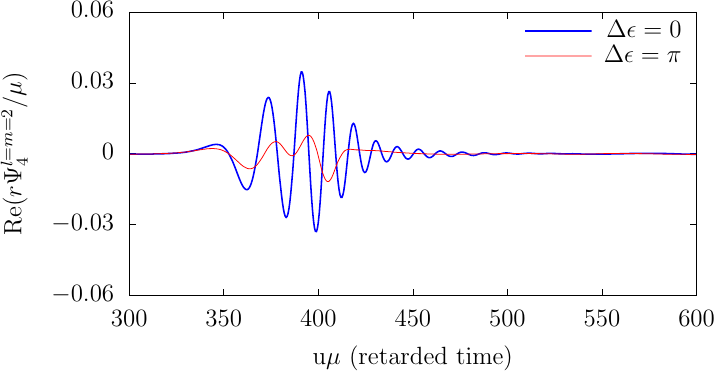}
} 
\caption{$l=m=2$ modes of $r\Psi_4^{lm}$ comparing the periodicity of the relative phase ($\Delta\epsilon$) for simulation $85$C$0$ and $85$C$6$ (top panel) and $85$C$3$ and $85$C$9$ (bottom panel) equal-mass merger ($\omega/\mu=0.8500$) comparing for periodic relative phase ($\Delta\epsilon$) for initial boost of $0.030$.}
\label{fig:GW_v015_omega08500_periodicity}
\end{figure}

\begin{figure*}[t]
  \centering
  \subfloat{
  \centering
  \includegraphics[width=0.5\linewidth]{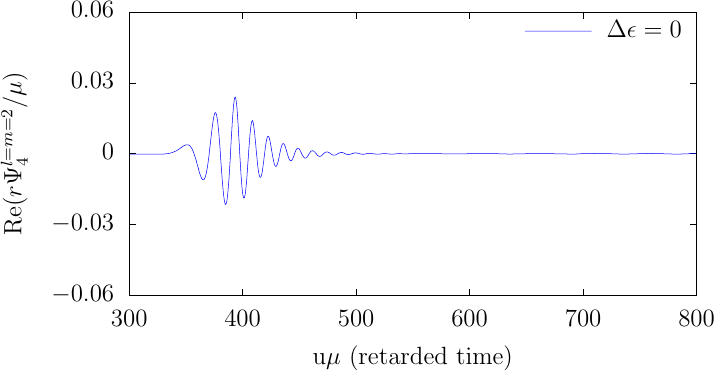}
}
  \subfloat{
  \centering
  \includegraphics[width=0.5\linewidth]{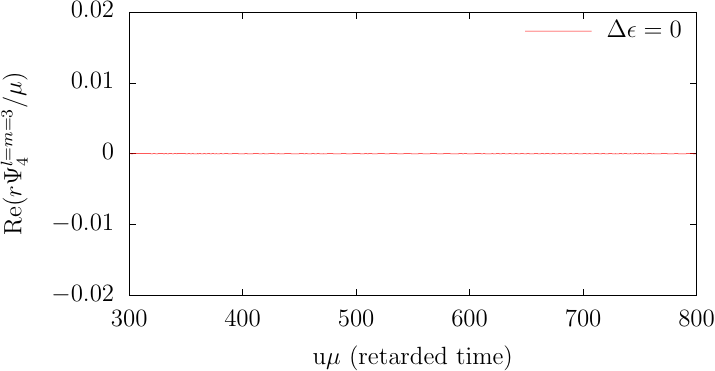}
}\\ 
  \subfloat{
  \centering
  \includegraphics[width=0.5\linewidth]{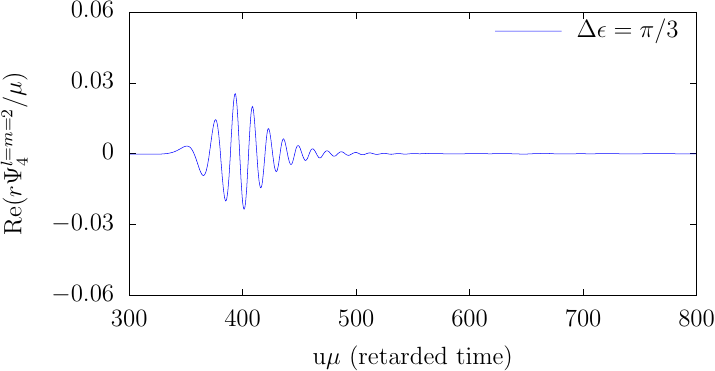}
}
  \subfloat{
  \centering
  \includegraphics[width=0.5\linewidth]{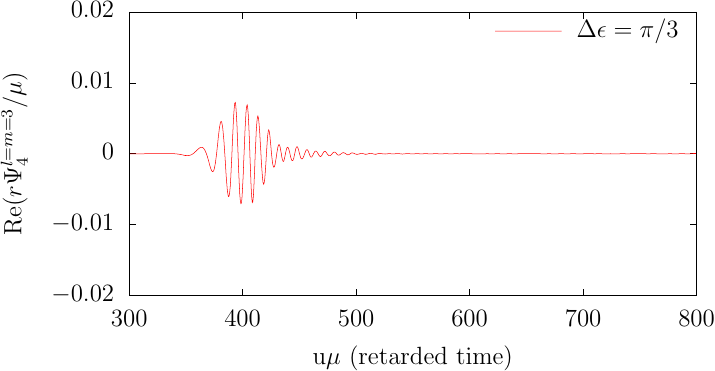}
}\\ 
  \subfloat{
  \centering
  \includegraphics[width=0.5\linewidth]{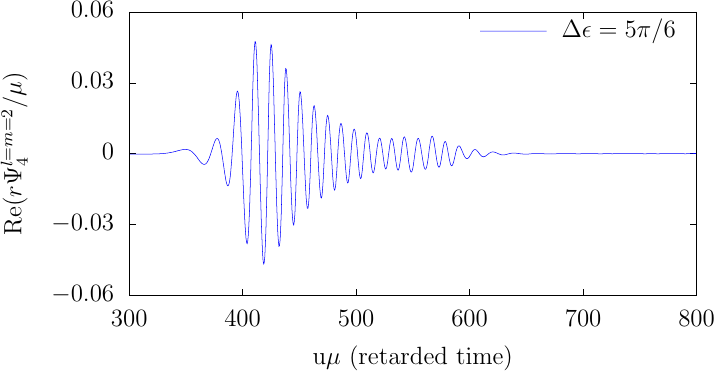}
}
  \subfloat{
  \centering
  \includegraphics[width=0.5\linewidth]{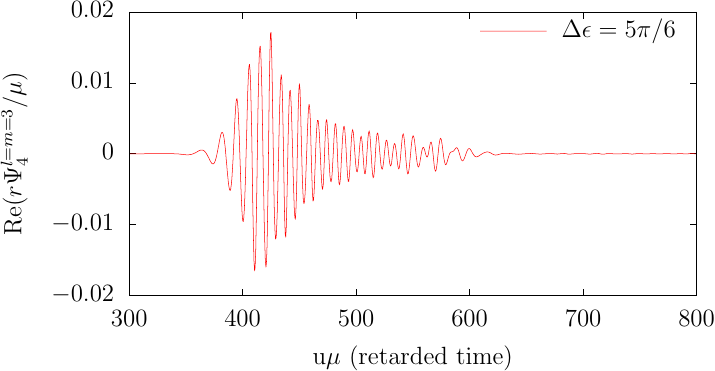}
}\\ 
  \subfloat{
  \centering
  \includegraphics[width=0.5\linewidth]{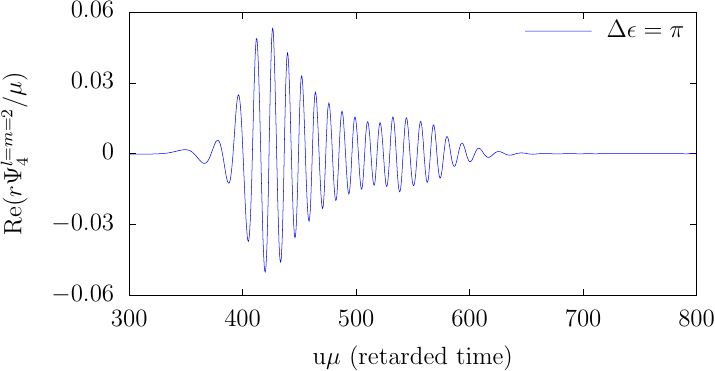}
}
  \subfloat{
  \centering
  \includegraphics[width=0.5\linewidth]{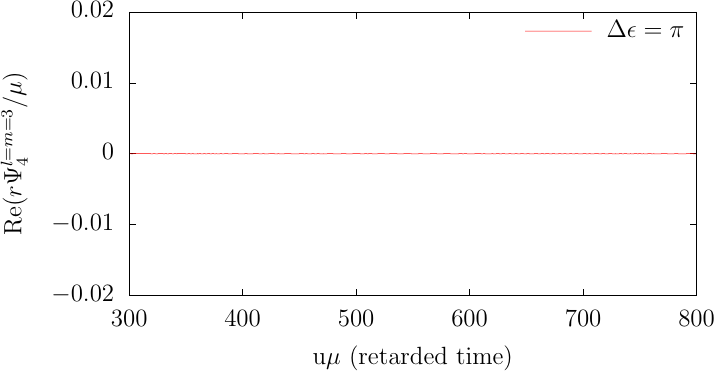}
}
\caption{Left column: $r\Psi_4^{22}$ waveforms for equal-mass mergers with $\omega/\mu=0.87$ and initial boost $v/c=0.020$ (models $87$B$0$, $87$B$2$, $87$B$5$ and $87$B$6$). Right column: $r\Psi_4^{33}$ waveforms for the same set of models.}
\label{fig:GW_lm22_lm33}
\end{figure*}

%%%%%%%%%%%%%%%%%%%%%%%%%%%%%%%%%%%%%%%%
\subsection{The role of the compactness}
%%%%%%%%%%%%%%%%%%%%%%%%%%%%%%%%%%%%%%%%

The compactness of the Proca stars also affects the dynamics and GW emision of the merger and post-merger phases. On the one hand, a larger compactness leads to the rapid collapse of the resulting object to a Kerr black hole surrounded, in some cases, by a short-lived Proca cloud~\cite{PhysRevD.75.064005}. On the other hand, less compact stars form transient hypermassive spinning Proca stars~\cite{PhysRevD.99.024017} or rotating bars, before collapsing to a Kerr black hole, delaying the time of collapse. It is worth noting that this behaviour replicates what is found in the case of binary neutron star mergers where more compact binaries lead to more rapid black hole formation (and larger GW emission)~\cite{Bernuzzi:2020}.

In terms of the effect of compactness on the GW emission, Figure \ref{fig:GW_comparev_e00} also reveals that the more compact stars show longer pre-collapse and post-merger phases of emission and larger values of the GW peak amplitude. The star's compactness, however, does not seem to affect the modes responsible for the emission. The dominant mode is always the quadrupolar one $l=m=2$ with subdominant contributions given by even-$m$ modes. When we take $\Delta\epsilon=0$, non-axisymmetric $(l,m)=(2,\pm 1)$ modes and, in general, odd-$m$ modes are completely suppressed due to the equal-mass symmetry of the system. We also note that the GW frequencies are hardly affected by compactness, only decreasing slightly as compactness increases.

%%%%%%%%%%%%%%%%%%%%%%%%%%%%%%%%%%%%%%%%%%%
\subsection{The role of the relative phase}
\label{rolephase}
%%%%%%%%%%%%%%%%%%%%%%%%%%%%%%%%%%%%%%%%%%%

Following~\cite{PhysRevD.75.064005,PhysRevD.106.124011,Siemonsen:2023a} we have also investigated the role of the initial relative phase on the merger dynamics and GW emission. Since the real and imaginary parts of the Proca field are both $\varphi$- and time-dependent, different initial phases correspond to different initial field orientations in $xy$ plane of the real and imaginary parts of the Proca field as shown in Figure~\ref{fig:chiphi_ep}. Such differences lead to distinct results revealing the internal phase structure of the stars. First, we check that the relative phase  $\Delta\epsilon=|\epsilon_1-\epsilon_2|$ is the key parameter that changes the outcome, independently of the values of the individual phases $\epsilon_1$ and $\epsilon_2$. To do so, we consider an equal-mass merger with $\omega/\mu=0.83$ and initial boost $v/c=0.015$, and vary the phase of both stars, always keeping the phase difference equal to zero ($\epsilon_1=\epsilon_2=\epsilon$ and $\Delta\epsilon=0$). Specifically, we choose three values of $\epsilon=\left\{0, \pi/4, \pi/2 \right\}$. Figure~\ref{fig:e00_v015_ep} shows the gravitational waveforms emitted by these systems, given by $r\Psi4^{\ell=m=2}$ extracted at $r_{\rm ext}=100$. We find that the waveforms are identical if the relative phase is kept the same. Correspondingly, in Figure \ref{fig:chiphi_ep} we exhibit a few snapshots of the time evolution of the energy density and the real part of the scalar potential $\chi_{\phi}$ for these cases. From the real part of the scalar potential, we confirm that, even though the initial orientation of the constituents of the Proca field is different due to the phase, there is no change in the evolution of the energy density. Therefore, the dynamics depends only on the relative phase and not on the individual phase structure of the star. Initially, the real (and imaginary) parts of the Proca field have a dipolar distribution in the $\varphi$ axis with $\bar m=1$ (top panels), but, after merger, a Proca remnant with a higher $\bar m$ mode can be seen around the black hole, in agreement with the results for head-on collisions  discussed in~\cite{PhysRevD.102.101504}. 

\begin{figure}
\centering
  \centering
  \includegraphics[width=1\linewidth]{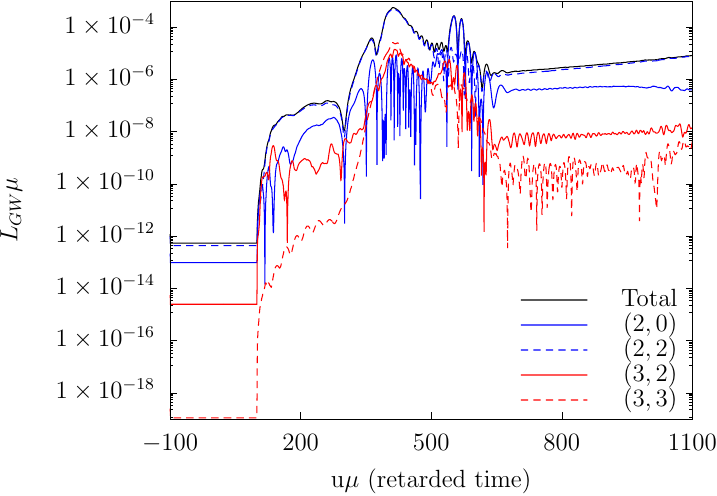}
\caption{Contribution to the total GW luminosity of several GW modes for model $87$B$5$. The dominant contribution is given by the $(l,m)=(2,2)$ quadrupolar mode.}
\label{fig:GW_energy_vs_t_modes}
\end{figure}

We now examine models with a non-zero initial relative phase, $\Delta\epsilon$. Although one might expect $\Delta \epsilon$ to take values spanning from 0 to $2\pi$ to cover the whole parameter space, it turns out that the results for $\Delta\epsilon$ and $2\pi-\Delta\epsilon$ are equivalent, as seen in the top panel of Figure \ref{fig:GW_v015_omega08500_periodicity}. Therefore, our simulations can be restricted to $\Delta\epsilon\in[0,\pi].$ Notably, the most pronounced differences in the waveforms occur for the in-phase ($\Delta\epsilon=0$) and anti-phase ($\Delta\epsilon=\pi$) binaries (see the bottom panel of Figure~\ref{fig:GW_v015_omega08500_periodicity}).

The phase has an important impact in the multipolar emission of the waveform~\cite{PhysRevD.106.124011}. Figure \ref{fig:GW_lm22_lm33} shows the real part of the $(l,m)=(2,2)$ (left column) and $(l,m)=(3,3)$ (right column) $\Psi_4$ modes for equal-mass binaries with $\omega/\mu=0.87$ and initial boost $v/c=0.02$. (Note that the waveforms are identical for $(l,m)=(2,-2)$ and $(l,m)=(3,-3)$, modulo a minus sign). For equal-mass Proca stars in-phase ($\Delta\epsilon=0$) or in anti-phase ($\Delta\epsilon=\pi$) the odd-$m$ modes are completely suppressed, although we will see later that under certain conditions, this statement does not hold entirely. This was also found in the equal-mass case in~\cite{PhysRevD.106.124011}. However, other values of the relative phase break the symmetry of the system and odd $(l,m)$ modes are triggered, attaining a significant contribution. In the case of head-on collisions~\cite{PhysRevD.106.124011, Bustillo:2023} the $(l,m)=(3,\pm 3)$ modes are triggered for both  equal and unequal-mass models, reaching, however, significantly lower amplitudes than in the eccentric models reported here. We observe that the amplitude of the $(l,m)=(3,\pm 3)$ mode increases with the relative phase, reaching its maximum for $\Delta \epsilon=5\pi/6$ and dropping to zero for $\Delta\epsilon=\pi$. We conjecture that an observation of large-amplitude odd modes in GW events from equal-mass or nearly equal-mass systems would serve as a smoking-gun indication of a non-trivial relative phase difference between the binary components and a signature of exotic physics, since equal-mass BBH mergers do not produce such modes (at least without precession).

Since eccentric mergers take longer to merge than head-on collisions, the overall radiated energy is also higher. In addition, the peak emission is also almost an order of magnitude larger. In either case, the emission is dominated by the quadrupolar modes. In order to understand the subdominant role of the odd-$m$ modes we plot the contribution of some of those modes to the GW luminosity (cf.~Eq.~\ref{eq:LGW}) in Figure \ref{fig:GW_energy_vs_t_modes} for model $87$B$5$. This figure clearly shows that the main contribution to the total GW luminosity is due to the quadrupolar ($l=m=\pm 2$) mode. 

\begin{figure}
  \centering
%  \subfloat{
%  \centering
%  \includegraphics[width=1.0\linewidth]{Plot/GW_e00_v020_omega083_bis.pdf}
%  \includegraphics[width=1.0\linewidth]{Plot/GW_e00_e06_v020_omega083_bis_new.pdf}
%} \\
  \subfloat{
  \centering
  \includegraphics[width=1.0\linewidth]{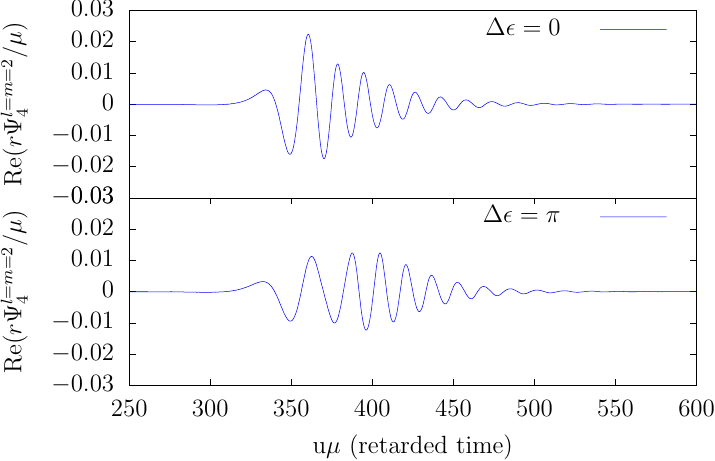}
}
\caption{Evolution of $r\Psi_4^{22}$ for models $83$B$0$ (in-phase on top panel) and $83$B$6$ (in-phase on bottom panel), both with $\omega/\mu=0.83$.} 
\label{fig:GW_ee_omega083}
\end{figure}

\begin{figure}[t]
\centering
\includegraphics[width=1.0\linewidth]{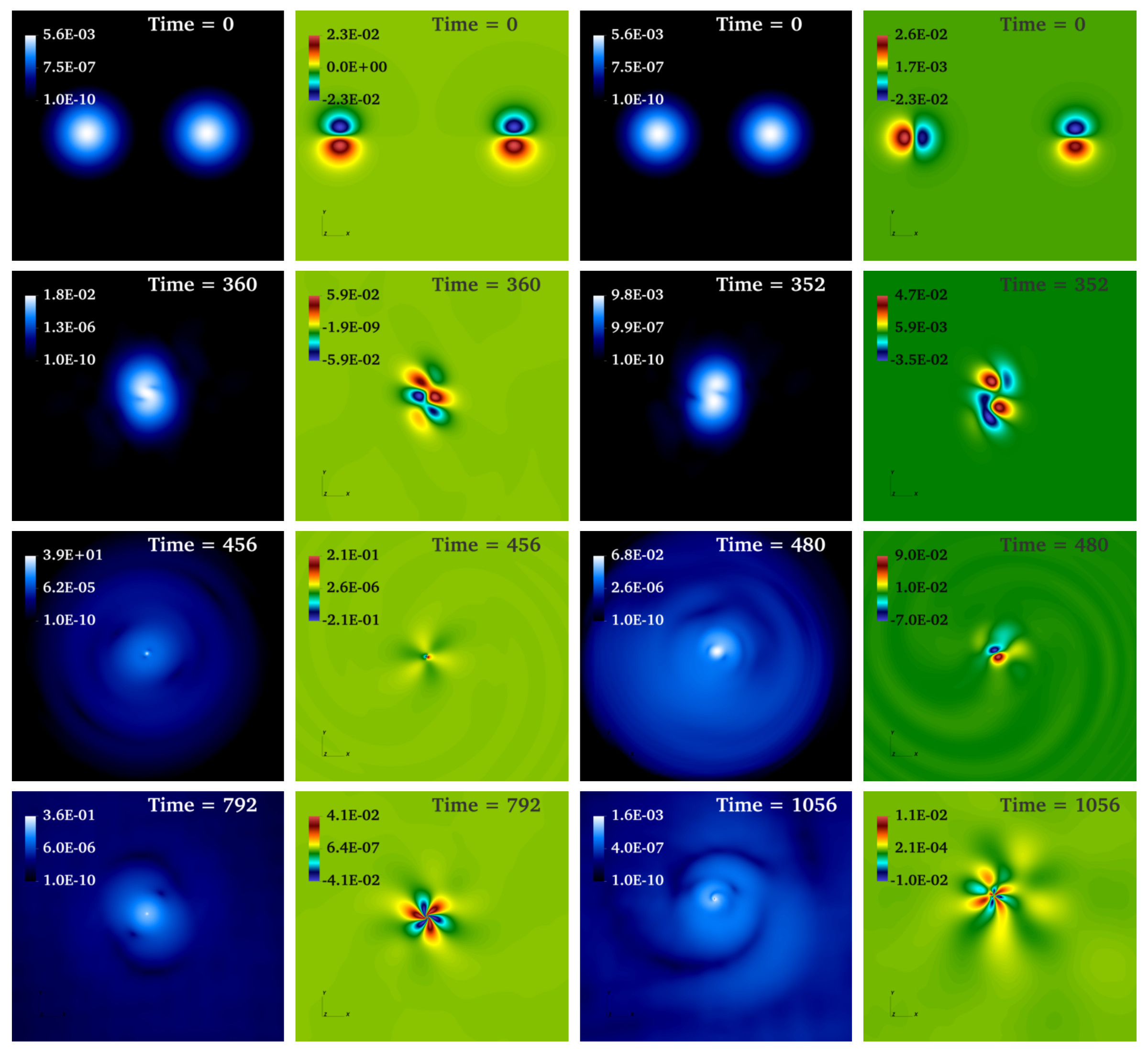}
\caption{Equatorial plane snapshots of the energy density (first and third columns) and of the real part of the scalar potential (second and fourth columns) for models $87$C$0$ (leftmost columns) and $87$C$3$ (rightmost columns).}
\label{fig:Snapshots_v030_e00_03_omega087}
\end{figure}

\begin{figure}[t]
\centering
\includegraphics[width=1.0\linewidth]{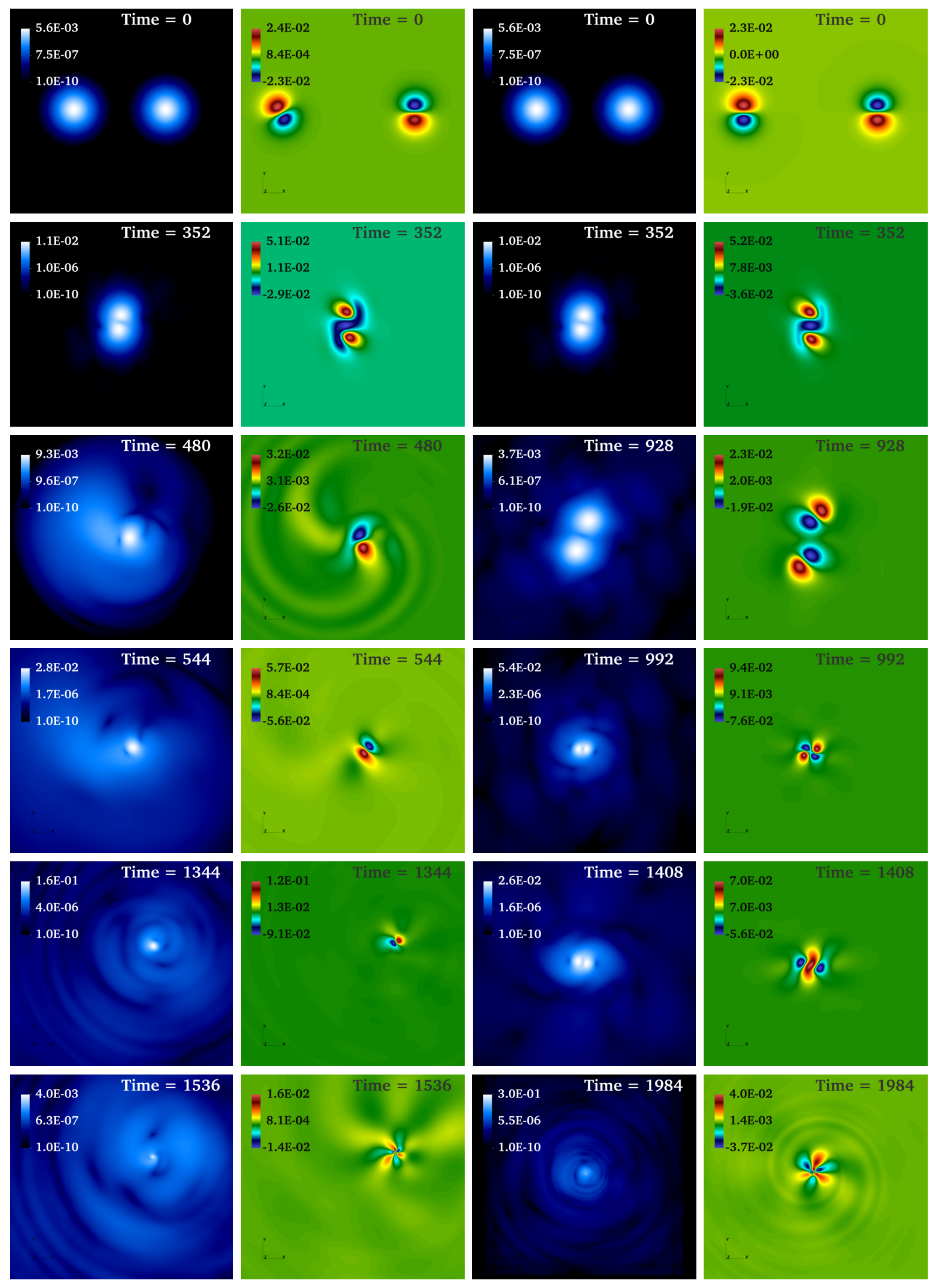}
\caption{Same as Fig.~\ref{fig:Snapshots_v030_e00_03_omega087} but for models $87$C$5$ (leftmost columns) and $87$C$6$ (rightmost columns).}
\label{fig:Snapshots_v030_e05_06_omega087}
\end{figure}

\begin{figure}
  \centering
  \subfloat{
  \centering
  \includegraphics[width=1\linewidth]{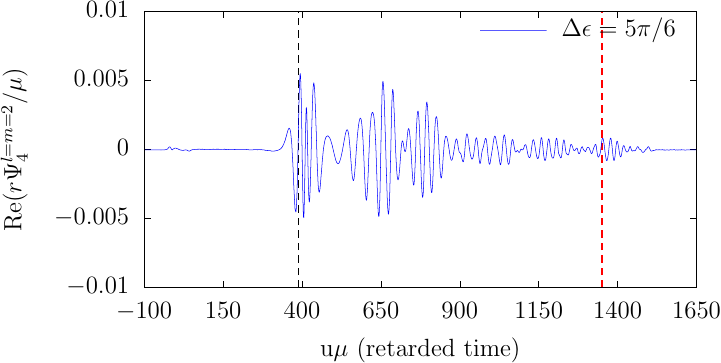}
}
\\
\subfloat{
  \centering
  \includegraphics[width=1\linewidth]{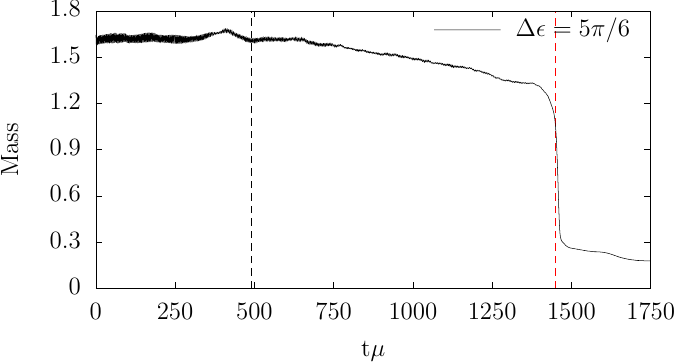}
}
\caption{Formation of a hypermassive Proca star for model $87$C$5$. Top panel: Evolution of $r\Psi_4^{22}$.
Bottom panel: Evolution of the integrated mass on the entire spherical volume of the computational grid. The dashed black line indicates the merger time ($\rm u \mu\sim390$), while the dashed red line corresponds to the time of black hole formation  ($\rm u \mu\sim1350$).}
\label{fig:e05_v030_omega087}
\end{figure}

\begin{figure*}
  \centering
  \subfloat{
  \centering
  \includegraphics[width=0.5\linewidth]{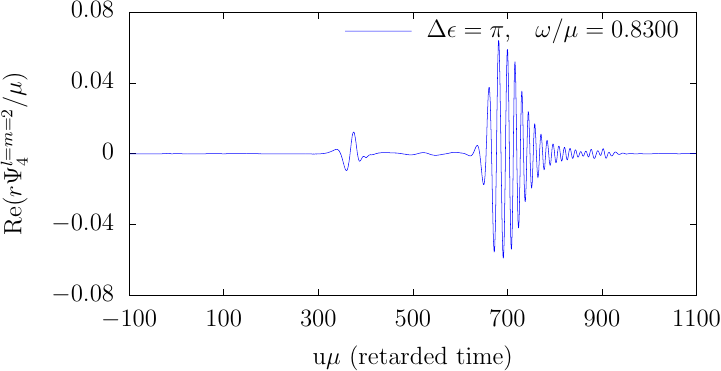}
}
  \subfloat{
  \centering
  \includegraphics[width=0.5\linewidth]{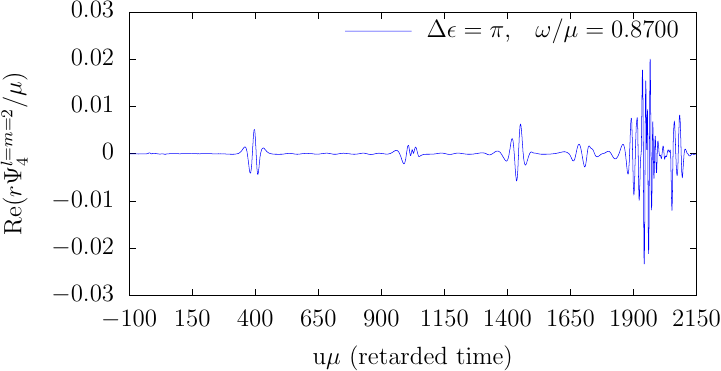}
}
\caption{Eccentric anti-phase models may yield more than one encounter. Evolution of $r\Psi_4^{22}$ for models $83$C$6$ (left) and $87$C$6$ (right). See also Fig.~\ref{fig:trajectories_BBS} for an illustration of the trajectories.}
\label{fig:GW_e06_v030_omega83_87}
\end{figure*}

\begin{figure}
\centering
\subfloat{
  \centering
  \includegraphics[width=1\linewidth]{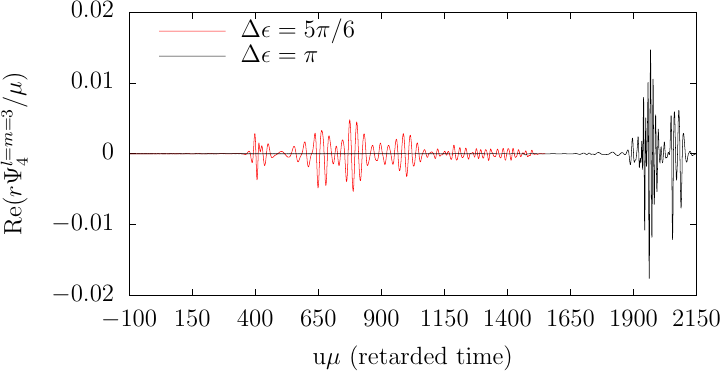}
}
\caption{$r\Psi_4^{33}$ GW modes for models $87$C$5$ (red solid line) and $87$C$6$ (black solid line).}
\label{fig:GW_m3_e05_06_v030_omega087}
\end{figure}

\begin{figure}
\centering
\subfloat{
  \centering
  \includegraphics[width=1\linewidth]{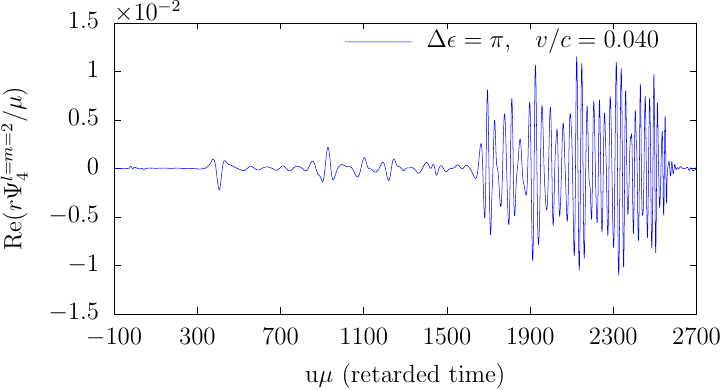}
}\\ 
\subfloat{
  \centering
  \includegraphics[width=1\linewidth]{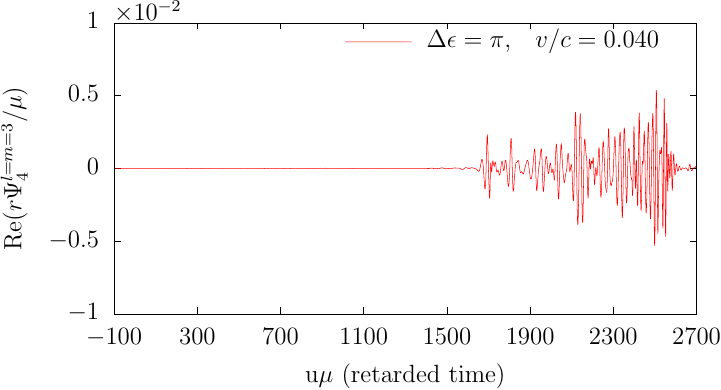}
}\\ 
\subfloat{
  \centering
  \includegraphics[width=1\linewidth]{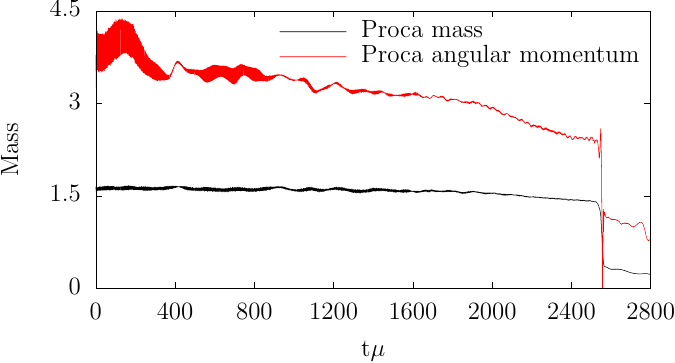}
}
\caption{$l=m=2$ (top panel) and $l=m=3$  (middle panel) modes of $r\Psi_4^{lm}$ and angular momentum and integrated Proca mass on spherical volume of radius equals to the radial extend of the grid (bottom panel) as a function of retarded time for simulation $87$D$6$.} 
\label{fig:GW_e06_v040_omega087}
\end{figure}

The morphology of the dominant $l=m=\pm 2$ mode is strongly affected by both the relative phase and by the compactness. The dependence on the relative phase is visible in the left column of Figure \ref{fig:GW_lm22_lm33} which shows that significantly longer signals are attained as the two stars become closer to be in anti-phase. This is due to the appearance of a a repulsive interaction between the two stars when they are in anti-phase that can keep them apart. In fact, in anti-phase one can even construct equilibrium configurations of two boson stars~\cite{Cunha:2022tvk,Ildefonso:2023qty} or even of two black holes with bosonic hair~\cite{Herdeiro:2023roz}. 
%\cite{PhysRevD.75.064005,PhysRevD.77.044036,ildefonso2023self}. 
For the case of boson star mergers the effects of an anti-phase were discussed in \cite{PhysRevD.75.064005,PhysRevD.77.044036,PhysRevD.95.124005,ildefonso2023self}. In head-on boson star collisions, the stars experience a repulsive force that surpasses gravitational attraction, preventing them from merging. As a result, the gravitational wave emission differs from that of zero-dephased boson stars. The total energy density of the system is given by $\rho =\rho_1+\rho_2+\Delta$, where $\Delta$ represents the effective interaction potential, which vanishes when the stars are well separated. It can be shown that $\Delta\propto \cos{\left(\Delta\epsilon\right)}$, meaning that for anti-phase boson stars, the gravitational potential develops a central bump with two surrounding minima. Therefore, merging is not energetically favorable for anti-phase boson stars. We observe similar results in the case of eccentric mergers, as discussed below. %\TF{This whole paragraph still does not read too well.}

Likewise, the amplitude of the quadrupolar mode also shows a monotonic dependence with the relative phase.
However, it follows an inverse dependence with the compactness. As depicted in Figure \ref{fig:GW_ee_omega083} the amplitude decreases as the stars are more compact. In particular for the less compact star ($\omega/\mu=0.87$) the time of merger (and black hole formation) and of the phase when the surrounding cloud is accreting are the longest (we refer the reader again to the left column of Fig.~\ref{fig:GW_lm22_lm33}, bottom panel). 

%\TF{From here on we start discussing again the role of the angular momentum. Probably this should have been discussed before, in Section A, and not in this Section C where we are discussing the role of the relative phase. However, since it is difficult to separate the two effects, I have decided to create a new section D where both effects are considered jointly. Feel free to improve the organization of the presentation.}

%%%%%%%%%%%%%%%%%%%%%%%%%%%%%%%%%%%%%%%%%%%%%%%%%%%%%%%%%%%%%%%%%%%%%%
\subsection{The joint role of the orbital angular momentum and relative phase}
%%%%%%%%%%%%%%%%%%%%%%%%%%%%%%%%%%%%%%%%%%%%%%%%%%%%%%%%%%%%%%%%%%%%%%

In Section~\ref{roleangularmomentum} and Section~\ref{rolephase} we analyzed, respectively, the isolated effects on the dynamics and on the waveforms of the orbital angular momentum (considering only in-phase systems) and of the relative phase (fixing the orbital angular momentum). Here we discuss the combined effect that varying both parameters has on the waveforms. 

We begin by considering the less compact configuration with $\omega/\mu=0.87$ and the highest initial orbital angular momentum ($v/c=0.030$). For this configuration, we explore models with varying relative phases. In Figure~\ref{fig:Snapshots_v030_e00_03_omega087}, the odd-numbered columns show snapshots of the energy density on the equatorial plane, while the even-numbered columns display the real part of the scalar potential. These panels correspond to models 87C0 (leftmost columns) and 87C3 (rightmost columns). Similarly, Fig.~\ref{fig:Snapshots_v030_e05_06_omega087} displays the same quantities for models 87C5 (leftmost columns) and 87C6 (rightmost columns). As both figures show, the internal structure of the stars, controlled by the relative phase, becomes relevant as the stars approach each other, leading to markedly different merger and post-merger evolutions. This effect is especially evident in the evolution of the scalar potential. Around the time of  merger ($\rm t\mu\sim360$, second row in both Fig.~\ref{fig:Snapshots_v030_e00_03_omega087} and Fig.~\ref{fig:Snapshots_v030_e05_06_omega087}), the energy density distributions look fairly similar but the scalar potential differ substantially.

For models 87C0 and 87C3, the merger leads to a compact object still characterized by the $\bar m=1$ scalar potential distribution (third row of Fig.~\ref{fig:Snapshots_v030_e00_03_omega087}) which rapidly collapses gravitationally into a black hole. For model 87C5, a more compact spinning $\bar m=1$ star forms. This star persists for up to $\rm t\mu\sim1500$ and experiences a noticeable kick before collapsing to a black hole. In contrast, for model 87C6, the repulsive interaction due to the anti-phase structure dominates the dynamics, triggering several bounces (see third row in Fig.~\ref{fig:Snapshots_v030_e05_06_omega087}) before eventually merging and forming a spinning $\bar m=2$ star with a toroidal energy density distribution (fourth and fifth rows). The formation of a bosonic star with a higher azimuthal number $\bar m$ due to the relative phase was already discussed in~\cite{Siemonsen:2023a}. Our result confirms their findings, showing that spinning stars can indeed be formed in orbital or eccentric mergers as long as the two stars have a non-zero phase difference~\cite{PhysRevD.95.124005}. 

Therefore, the merger and post-merger dynamics is affected by the relative phase in a remarkable way: the system can either undergo prompt collapse to a black hole with an accreting Proca cloud (not in equilibrium), as in the case of models 87C0 and 87C3, or experience a prolonged transient phase corresponding to a highly-spinning ``hypermassive'' Proca star (model 87C5). This is a transient remnant which has an excess angular momentum that allows it to support more mass. Moreover, in the case of model 87C6 a $\bar m=2$ spinning Proca star is formed. However, such configurations are unstable against non-axisymmetric perturbations~\cite{PhysRevLett.123.221101,di2020dynamical,siemonsen2021stability}.

Figure \ref{fig:e05_v030_omega087} shows the $l=m=2$ waveform (top panel) for model $87$C$5$ discussed above along with the evolution of the integrated Proca mass on the full spherical volume of the grid. This evolution corresponds to the dynamics depicted in the leftmost column of Fig.~\ref{fig:Snapshots_v030_e05_06_omega087}. This waveform is strikingly different to all previous waveforms discussed in this work, both in terms of morphology and duration. The reason is because for this model the merger (represented by the black dashed line) does not lead promptly to a black hole (whose formation is represented by the red dashed line) but rather to the formation of a hypermassive Proca star. The bottom panel of Fig.~\ref{fig:e05_v030_omega087} shows that, at the verge of collapsing, the total mass of the star is $M\mu\sim1.3$, while the maximum mass of $\bar m =1$ spinning Proca stars is $M\mu_{\rm max}=1.125$. The extra angular momentum is radiated away by GW and the mass is also dispersed away through the gravitational cooling mechanism. Both processes trigger oscillations of the remnant and leave an imprint on the waveform. Once the excess angular momentum has been radiated away the hypermassive star can no longer support its additional mass and collapses to a black hole, indicated by the red dashed line in Fig.~\ref{fig:e05_v030_omega087}. Interestingly, the final black hole acquires a noticeable kick, visible in the energy density plot of Fig.~\ref{fig:Snapshots_v030_e05_06_omega087} at $\rm t\mu\sim1300$ (for model $87$C$5$). Similar results are obtained for other models of our sample, namely models $87$C$2$ and $87$C$4$, which also show the formation of transient hypermassive $\bar m=1$ Proca stars.

Let us now consider the merger of two anti-phase models, models $83$C$6$ and $87$C$6$, which differ by their compactness. Both start with the highest orbital angular momentum of our sample ($v/c=0.030$). Figure \ref{fig:GW_e06_v030_omega83_87} shows the quadrupole $\ell=m=2$ mode of the waveforms for these two models. The impact of the relative phase for mergers in anti-phase is significant. The combined effect of stars in anti-phase and with high angular momentum leads to situations in which there can be more than one encounter, particularly for less compact stars (model 87C6, right plot of Fig.~\ref{fig:GW_e06_v030_omega83_87}).   Each encounter produces a visible burst of GWs in the waveform. We recall that the trajectories of the two stars for model 87C6 are plotted on the right panel of Fig.~\ref{fig:trajectories_BBS}.
 
Another interesting effect is observed in these anti-phase models exhibiting multiple encounters. Our results show that the initial phase difference is conserved throughout the merger. This explains the multiple encounters for $\Delta\epsilon=\pi$, as the repulsive interaction keeps the stars separated for a longer time before they settle into a final configuration. Notably, while non-zero phase differences generally induce significant emission of the $\ell=m=3$ mode (see Fig.~\ref{fig:GW_lm22_lm33}), odd-$m$ modes are typically suppressed in both the in-phase ($\Delta\epsilon=0$) and anti-phase ($\Delta\epsilon=\pi$) configurations. When the initial orbital angular momentum is increased, the emission of the $\ell=m=3$ mode remains important after the collision. For instance, Figure \ref{fig:GW_m3_e05_06_v030_omega087} (red solid line) shows that the $\ell=m=3$ mode in model 87C5 reaches an amplitude nearly comparable to that of the quadrupole $\ell=m=2$ mode for $\Delta\epsilon=5\pi/6$. However, the black solid line in Fig.~\ref{fig:GW_m3_e05_06_v030_omega087} corresponds to the 87C6 model, revealing that it also emits a large $\ell=m=3$ mode despite the stars being in anti-phase. We observe that the $\ell=m=3$ mode is zero up to $\rm u\mu\sim1800$ and is simultaneously radiated with the peak of the $\ell=m=2$ mode (see right panel of Fig.~\ref{fig:GW_e06_v030_omega83_87}). This burst of GWs is not associated with the merger itself but rather with the non-axisymmetric instability of the $\bar m=2$ spinning Proca star. The instability breaks the anti-phase condition that initially allowed the formation of the $\bar m=2$ star, leading to a decay to a $\bar m=1$ spinning Proca star and then to the gravitational collapse to a black hole. These processes are accompanied by the GW emission of odd-modes.

\begin{figure}
  \centering
  \includegraphics[width=1.0\linewidth]{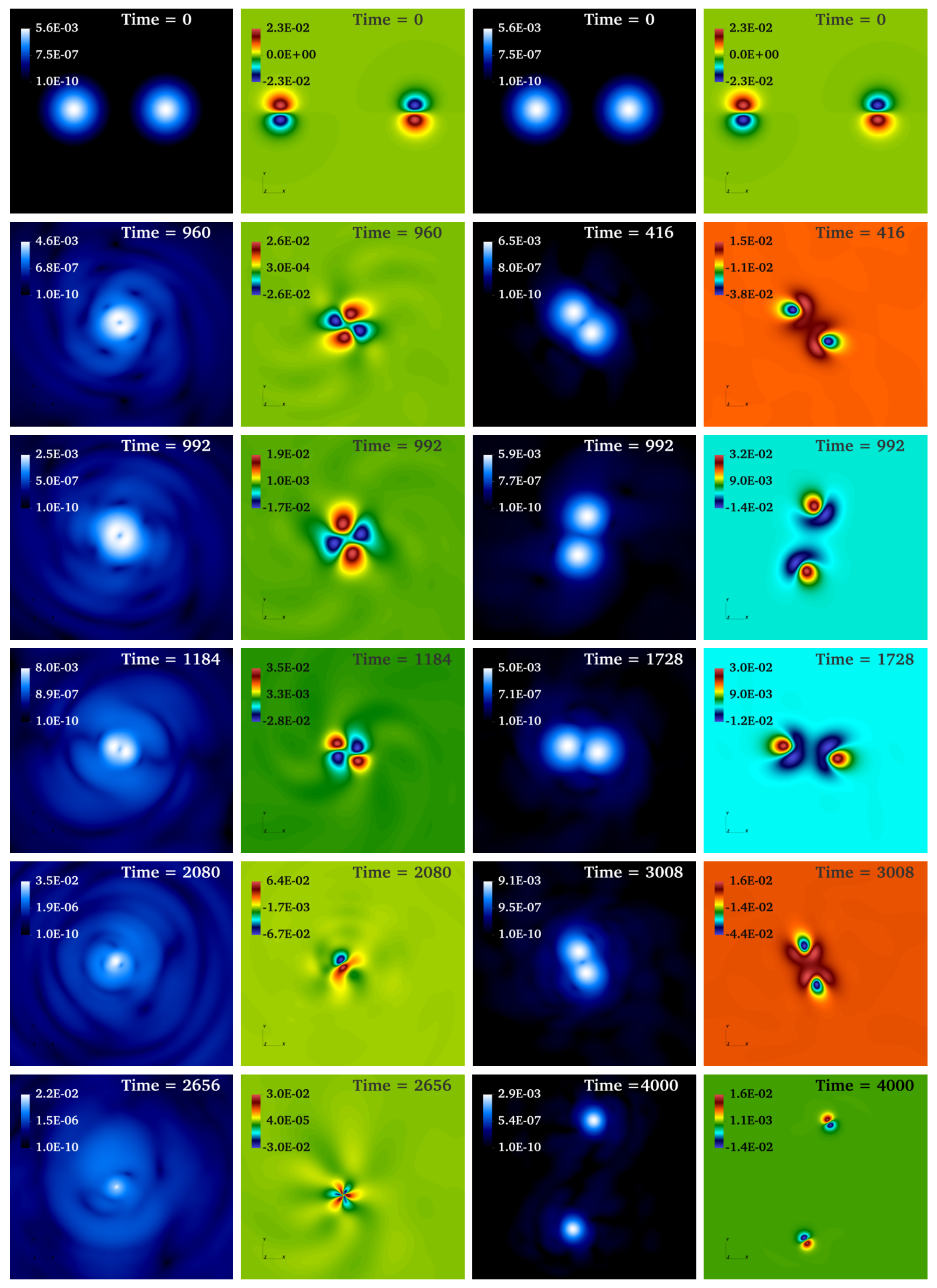}
\caption{Same as Fig.~\ref{fig:Snapshots_v030_e00_03_omega087} but for models $87$D$6$ (leftmost columns) and $87$E$6$ (rightmost columns).}
\label{fig:Snapshots_v040_050_e06_omega087}
\end{figure}

We further investigate the effect of the initial orbital angular momentum for merging Proca stars in anti-phase, considering two more cases (87D6 and 87E6) where the initial boost is even larger ($v/c=0.040$ and $v/c=0.050$). Figure~\ref{fig:GW_e06_v040_omega087} shows the $l=m=2$ (top panel) and $l=m=3$  (middle panel) waveforms for model $87$D$6$, along with the time evolution of the Proca mass and angular momentum (bottom panel). The dynamics for this model is plotted in the two leftmost columns of Figure~\ref{fig:Snapshots_v040_050_e06_omega087} which display snapshots of the real part of the scalar potential and of the corresponding energy density on the equatorial plane. In this case, a spinning $\bar m=2$ Proca star forms following merger. The energy density acquires a toroidal shape which is distinct from the spheroidal $\bar m=1$ configuration. It then becomes unstable to a non-axisymmetric instability~\cite{di2020dynamical}, loses angular momentum and mass, and decays to a hypermassive $\bar m=1$ around $\rm t\mu\sim 2000$. As we can see in the bottom panel of Fig.~\ref{fig:GW_e06_v040_omega087}, after the instability the mass of the new $\bar m=1$ star is around $M\mu\sim1.5$, which is larger than the maximum mass of spinning $\bar m=1$ Proca star. This leads to the eventual gravitational collapse and black hole formation at around $\rm u\mu\approx 2400$. Therefore, in the anti-phase cases, the $\ell=m=3$ GW mode is only triggered during the development of the non-axisymmetric instability of the unstable $\bar m=2$ Proca star, reaching almost the same order of magnitude of the quadrupolar mode emission.

\begin{figure}
\centering
\subfloat{
  \centering
  \includegraphics[width=1\linewidth]{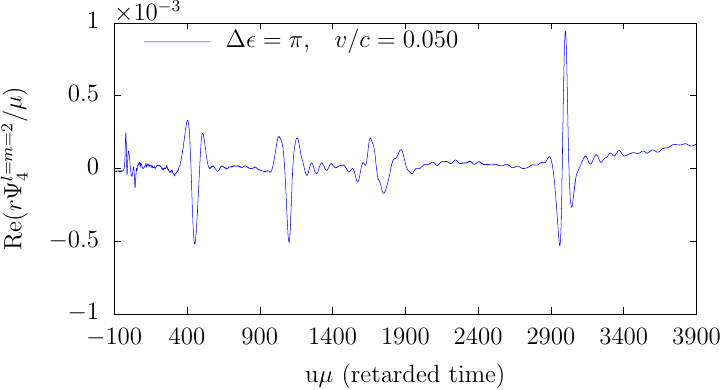}
}
\caption{$r\Psi_4^{22}$ waveform for model $87$E$6$.}
\label{fig:GW_e06_v050_omega087}
\end{figure}

If we increase the initial orbital angular momentum  even further (model $87$E$6$, $v/c=0.050$) the two stars do not merge and escape from each other in a hyperbolic encounter. This is visible in the snapshots of the energy density and of the real part of the scalar potential for this model shown in the two rightmost columns of Fig.~\ref{fig:Snapshots_v040_050_e06_omega087}. The corresponding quadrupole waveform for this model is shown in Figure \ref{fig:GW_e06_v050_omega087}. Since there is no merger and $\Delta\epsilon$ remains zero, the odd-$m$ modes are suppressed in this case. 

\begin{figure}[t]
\centering
\subfloat{
  \centering
  \includegraphics[width=1\linewidth]{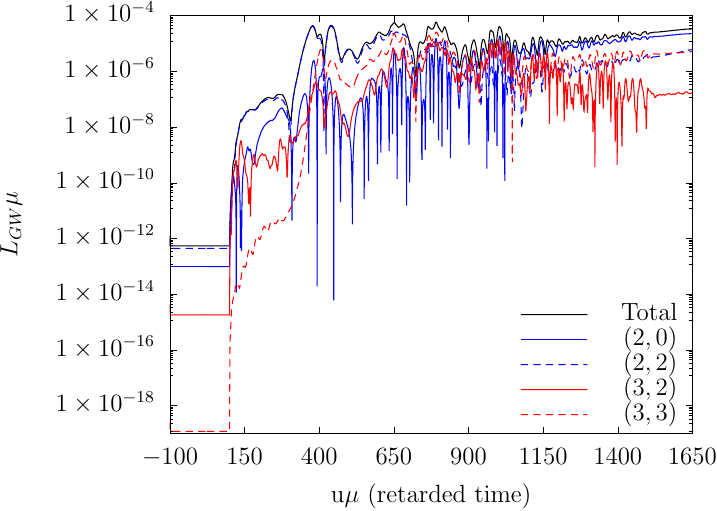}
}\\ 
\subfloat{
  \centering
  \includegraphics[width=1\linewidth]{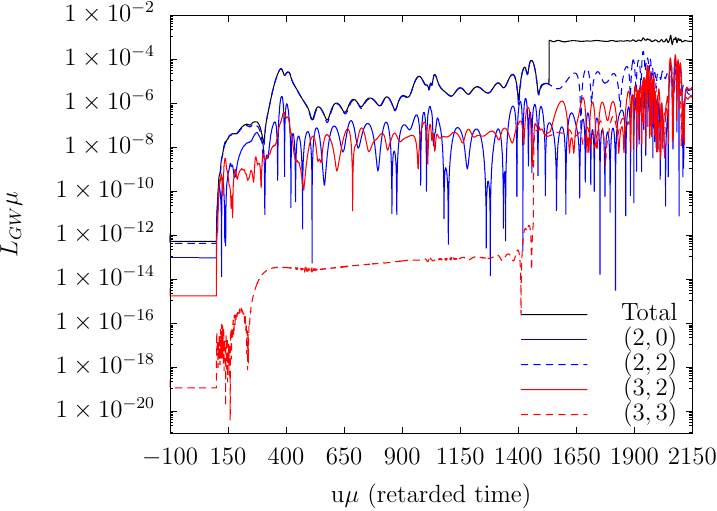}
}
\caption{Evolution of the GW luminosity associated with different emission modes for models $87$C$5$ (top panel) and $87$C$6$ (bottom panel). The total GW luminosity is displayed with black solid lines.}
\label{fig:LGW_m3_e05_06_v030_omega087}
\end{figure}

\begin{figure*}
\centering
\subfloat{
  \centering
  \includegraphics[width=0.32\linewidth]{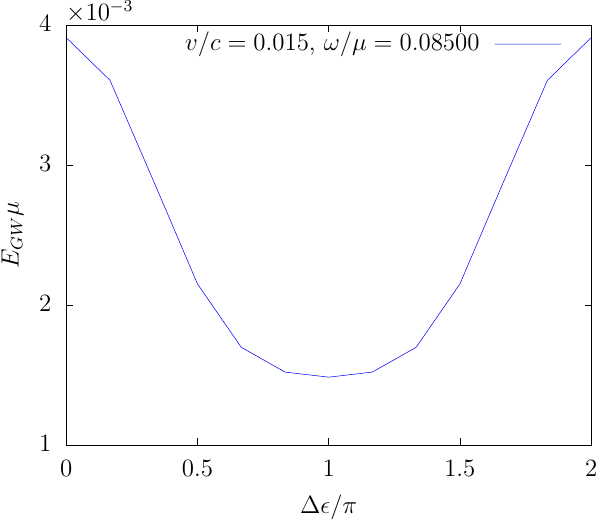}
}
\subfloat{
  \centering
  \includegraphics[width=0.32\linewidth]{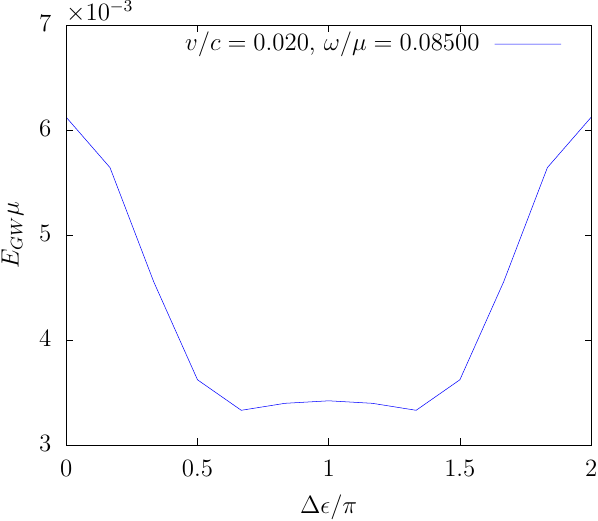}
}
\subfloat{
  \centering
  \includegraphics[width=0.32\linewidth]{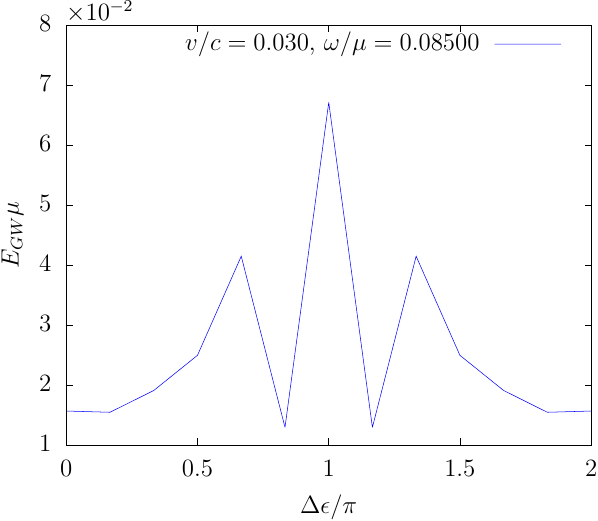}
}\\
\subfloat{
  \centering
  \includegraphics[width=0.32\linewidth]{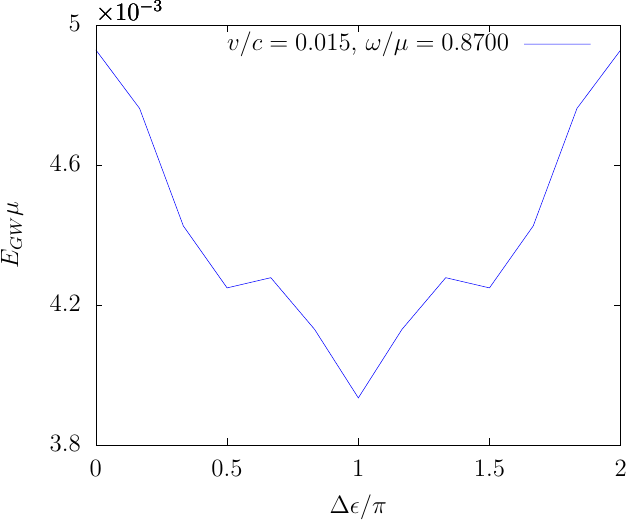}
}
\subfloat{
  \centering
  \includegraphics[width=0.32\linewidth]{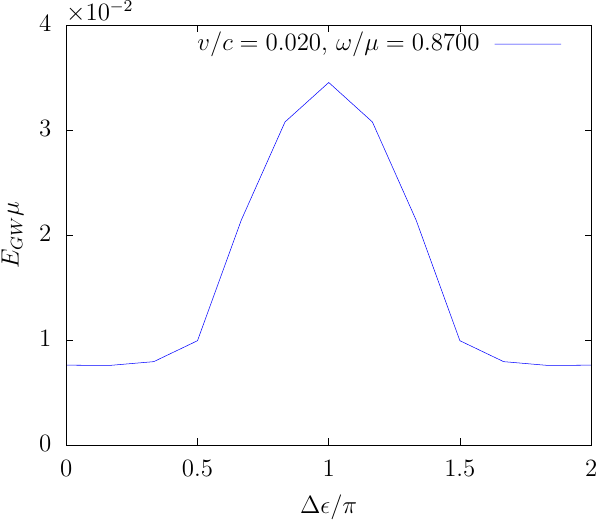}
}
\subfloat{
  \centering
  \includegraphics[width=0.32\linewidth]{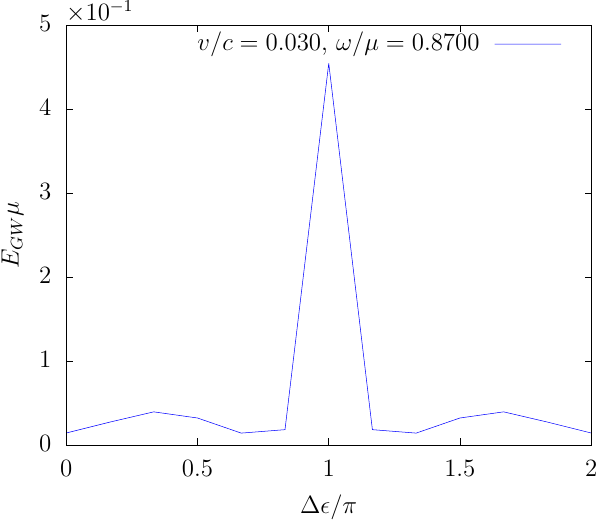}
}
\caption{Dependence of the GW energy on the relative phase for equal-mass Proca star mergers with $\omega/\mu=0.8500$ (top row) and $0.8700$ (bottom row), for $v/c=0.015$, $0.020$ and $0.030$ (from left to right). The relative phase is in units of $\pi$.}
\label{fig:GW_energy}
\end{figure*}

%%%%%%%%%%%%%%%%%%%%%%%%%%%%%%%%%%%%%
\subsection{GW energy and luminosity}
%%%%%%%%%%%%%%%%%%%%%%%%%%%%%%%%%%%%%

We end this study by analysing the energy emitted in gravitational waves in eccentric mergers of Proca stars. 
Fig.~\ref{fig:LGW_m3_e05_06_v030_omega087} compares the GW luminosity for various modes as well as the total emission, for models 87C5 and 87C6. As stated before, the contribution of the odd-$m$ modes is not subdominant for a relative phase difference of $\Delta\epsilon=5\pi/6$ and it becomes non negligible for anti-phase Proca star binaries (the peak luminosity is reached at the time of merger). 

The relative phase changes the pattern of the Proca field at the time of merger. It is possible to approximate the real and imaginary parts of the field such that the complex amplitude of the Proca field reads
\begin{equation}\label{eq:GW_em}
\begin{split}
    |\mathcal{A}|^2&=\text{Re}(\mathcal{A})^2+\text{Im}(\mathcal{A})^2\sim 4\cos^2\left[\frac{(\omega_1-\omega_2)}{2}t+\Delta\epsilon\right]=\\
    &=2\left\{1+\cos\left[(\omega_1-\omega_2)t+\frac{\Delta\epsilon}{2}\right]\right\}\,.
\end{split}
\end{equation}
The square of the amplitude of the Proca field is proportional to the GW energy, $E_{GW}\propto |\mathcal{A}|^2$.
We can use this analytic estimate to see how the relative phase structure affects the GW energy emission. This procedure must be regarded as an approximation since other factors such as the dynamics of the merger, the radius of the stars, and the time of merger, also affect the GW emission. Fig.~\ref{fig:GW_energy} displays the GW energy as a function of $\Delta\epsilon$ for models with $\omega/\mu=0.85$ and $0.87$ for the three different values of the initial boosts. The observed trend is similar to what we found in the case of head-on collisions of equal-mass Proca stars~\cite{PhysRevD.106.124011}, even though in the case of eccentric mergers there is no perfect destructive interference. While Eq.~(\ref{eq:GW_em}) describes a purely periodical trend in the GW energy as a function of the relative phase, Fig.~\ref{fig:GW_energy} shows that other profiles arise depending on the initial orbital angular momentum. In particular, in the head-on case, the minimum in the GW energy is attained for anti-phase Proca stars. This is no longer true when we increase the initial boost as, in this case, additional features appear in the dynamics (repulsive interaction, multiple encounters, formation of unstable $\bar m=2$ Proca stars and gravitational collapse). These changes in the GW energy ultimately reflect how the merger dynamics responds to different interior phase structures in the stars. The comparison between the least luminous model ($85$A$6$) and the most luminous one (model $87$C$6$) shows that the latter is almost $300$ times brighter. 

\begin{figure}
\centering
\subfloat{
  \centering
  \includegraphics[width=1\linewidth]{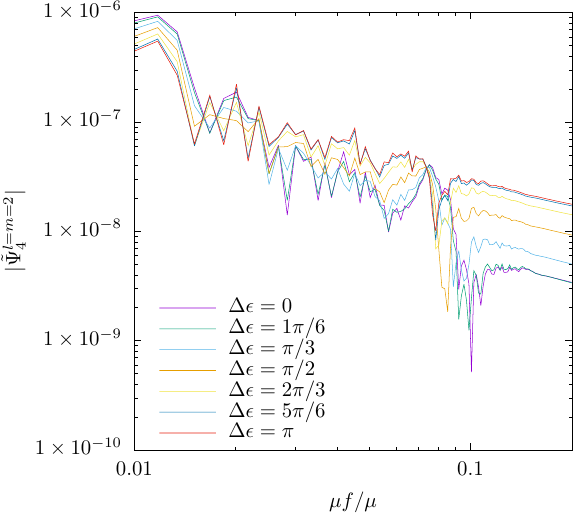}
}
\caption{Absolute value of the Fourier transform of the $l=m=2$ mode of $\Psi_4$ for equal-mass PSs orbital merger with $\omega/\mu=0.8300$ and initial boost $v/c=0.020$ as a function of the relative phase of the two stars.}
\label{fig:FFT_083_v020}
\end{figure}

The role of the relative phase $\Delta\epsilon$ also influences the structure of the emission modes and their frequency content (their morphology). For Kerr black holes, the ringdown is dominated by the quadrupolar $(l,m)=(2,2)$ mode while all other multipoles are subdominant~\cite{zhao2022quasinormalmodesblack}. Fig.~\ref{fig:FFT_083_v020} shows the Fourier transform of the quadrupole mode for model 83BX ($\omega/\mu=0.83$ and $v/c=0.02$) as a function of the relative phase $\Delta \epsilon$. The amplitude of the mode and its frequency content are significantly modified by the interior phase structure. These changes could be large enough to be measurable in a Bayesian model selection framework. Therefore, the observed dependency could provide a possible means to distinguish bosonic star mergers from other types of mergers of compact objects (in particular BBH mergers). 

%This would impact on the observability of the source, suggesting that this effect could actually  be measurable in a Bayesian parameter interference framework. This dependency  has a potential impact in GW  data  analysis being a possible indicator to distinguish bosonic star mergers from other compact objects. 

%Finally, Figure~\ref{fig:psi33_083_v020} shows how the odd-$m$ mode are triggered when  $\Delta \epsilon \ne 0$ and $\pi$. \NS{This figure could be moved to an earlier discussion or removed}

%\begin{figure}
%\subfloat{
%  \centering
%  \includegraphics[width=1\linewidth]{Plot/Psi33module_v020_omega083_bis.pdf}
%}
%\caption{Absolute value of $l=m=3$ mode of $\Psi_4$ for equal-mass PSs orbital merger with $\omega/\mu=0.8300$ and initial boost $v/c=0.020$ as a function of the relative phase of the two stars: the odd-$m$ modes are activated when $\Delta \epsilon \ne 0$ and $\pi$.}
%\label{fig:psi33_083_v020}
%\end{figure}

%%%%%%%%%%%%%%%%%%%%%
\section{Conclusions}
\label{conclusions}
%%%%%%%%%%%%%%%%%%%%%

In this work we have performed a comprehensive study of eccentric mergers of spinning Proca stars. The conclusions of our study are drawn after the analysis of a dataset that comprises about 100 numerical relativity simulations of such mergers. Those simulations have allowed us to explore the impact of the orbital angular momentum, compactness, and relative phase structure on the mergers' outcome. Our results have shown that the internal phase structure of Proca stars, characterized by the relative phase difference $\Delta\epsilon$, remains unchanged during the merger and critically influences the dynamics and GW emission. In anti-phase configurations ($\Delta\epsilon = \pi$), the appearance of a repulsive interaction can lead to multiple encounters before the final merger, prolonging the interaction and producing distinct GW bursts. Non-zero relative phase differences (other than $\Delta\epsilon = 0$ or $\pi$) can trigger odd-$m$ GW modes with significant amplitudes (e.g., $\ell = m = 3$), even when odd-$m$ modes are typically suppressed in strictly in-phase or anti-phase cases.  Increasing the initial orbital angular momentum (via higher boost velocities, e.g., $v/c = 0.030$) enhances the GW emission, making odd-$m$ modes prominent and affecting the duration of the post-merger phase.

Different merger outcomes have been observed depending on the combination of orbital angular momentum and relative phase. Some of our models (e.g.~87C0 and 87C3) form a compact object with an $\bar m = 1$ scalar potential that quickly collapses to a black hole. Model 87C5 produces a more compact spinning $\bar m = 1$ star that survives much longer and even experiences a recoil kick before collapsing. Model 87C6 shows that the anti-phase repulsion can trigger multiple bounces, eventually leading to the formation of a spinning $\bar m = 2$ star with a toroidal energy density distribution, in agreement with previous results in the scalar case.

We argue that the unique GW signatures observed in this investigation, especially the emergence of large-amplitude odd-$m$ modes, could serve as observational proofs of exotic physics, distinguishing these mergers from conventional equal-mass BBH mergers. Moreover, the gravitational waveforms computed in the numerical simulations reported in this work enlarge the waveform catalog of head-on collisions of Proca stars reported in~\cite{PhysRevD.106.124011}. Our study provides new waveform templates against which match-filter the data collected through modeled searches of mergers of exotic compact objects. Given the observational capabilities of present day GW detectors and those expected for third-generation observatories, the production of waveform catalogs of physically motivated exotic compact objects as the ones investigated in this work, is a well-timed, worth-pursuing effort. 

%%%%%%%%%%%%%%%%%%%%%%%%%%%%%%%%%%%%%%%%%%%%%%%%%%%%%%%%%%%%%%%%%%%%%%%%%%%%%%
\begin{acknowledgments}
%%%%%%%%%%%%%%%%%%%%%%%%%%%%%%%%%%%%%%%%%%%%%%%%%%%%%%%%%%%%%%%%%%%%%%%%%%%%%%
%
GP acknowledges support from the Spanish Agencia Estatal de Investigaci\'on through grant PRE2022-104185
funded by MICIU/AEI/10.13039/501100011033 and by FSE+. NSG acknowledges support from the Spanish Ministry of Science and Innovation via the Ram\'on y Cajal programme (grant RYC2022-037424-I), funded by MCIN/AEI/10.13039/501100011033 and by ``ESF Investing in your future”. GP, NSG, and JAF are supported by the Spanish Agencia Estatal de Investigaci\'on (grant PID2021-125485NB-C21) funded by MCIN/AEI/10.13039/501100011033 and ERDF A way of making Europe, and by the Generalitat Valenciana (grant CIPROM/2022/49). We acknowledge further support from the European Horizon Europe staff exchange (SE) programme HORIZON-MSCA2021-SE-01 Grant No. NewFunFiCO-101086251 and from the Center for Research and Development in Mathematics and Applications (CIDMA) through the Portuguese Foundation for Science and Technology (FCT -- Fundaç\~ao para a Ci\^encia e a Tecnologia) under the Multi-Annual Financing Program for R\&D Units, PTDC/FIS-AST/3041/2020 (\url{http://doi.org/10.54499/PTDC/FIS-AST/3041/2020}),  2022.04560.PTDC (\url{https://doi.org/10.54499/2022.04560.PTDC}) and 2024.05617.CERN (\url{https://doi.org/10.54499/2024.05617.CERN}). 
The authors acknowledge the computer resources provided by the Red Espa\~nola de Supercomputaci\'on (Tirant, MareNostrum5, and Storage5) and the technical support from the IT departments of the Universitat de Val\`encia and the Barcelona Supercomputing Center (allocation grants RES-FI-2023-1-0023, RES-FI-2023-2-0002, RES-FI-2024-2-0012, and RES-FI-2024-3-0007). 

%This manuscript has LIGO DCC number XXXX.

%%%%%%%%%%%%%%%%%%%%%%%%%%%%%%%%%%%%%%%%%%%%%%%%%%%%%%%%%%%%%%%%%%%%%%%%%%%%%%
\end{acknowledgments}
%%%%%%%%%%%%%%%%%%%%%%%%%%%%%%%%%%%%%%%%%%%%%%%%%%%%%%%%%%%%%%%%%%%%%%%%%%%%%%

\bibliography{Biblio}

\end{document}